\documentclass[aps,prl,notitlepage,superscriptaddress,twocolumn,amsmath,amssymb,10pt]{revtex4-1}
\usepackage{times,bbm}

\usepackage[usenames,dvipsnames]{xcolor}
\usepackage{mathtools}
\usepackage{tikz}  
\usetikzlibrary{arrows,shapes,positioning,shadows,backgrounds,fit}
\usepackage[normalem]{ulem}

\usepackage[english]{babel}
\usepackage{graphicx, dcolumn, bm, color, amsmath,amsthm, relsize, amsmath, dsfont, mathrsfs, empheq, verbatim, upgreek, etoolbox,braket}
\usepackage[caption = false]{subfig}

\usepackage[colorlinks=true,linkcolor=blue,urlcolor=blue,citecolor=blue]{hyperref}
\usepackage{graphics}
\usepackage{bm}
\usepackage{color}
\usepackage{amscd}
\usepackage{amsfonts}
\usepackage{amssymb}
\usepackage{graphicx}
\usepackage{tabularx}
\usepackage{tikz}
\usetikzlibrary{quantikz}

\usepackage{amsthm}
\theoremstyle{definition}

\newtheorem*{theorem*}{Theorem}

\newtheorem*{corollary*}{Corollary}

\newtheorem*{lemma*}{Lemma}

\newdimen\origiwspc%
\newdimen\origiwstr%
\preprint{}

\begin{document}

\title{Evidence of Kardar-Parisi-Zhang scaling on a digital quantum simulator}

\author{Nathan Keenan}
\email{nakeenan@tcd.ie}
\affiliation{IBM Quantum, IBM Research Europe - Dublin, IBM Technology Campus, Dublin 15, Ireland}
\affiliation{Department of Physics, Trinity College Dublin, Dublin 2, Ireland}

\author{Niall F. Robertson}
\email{niall.robertson@ibm.com}
\affiliation{IBM Quantum, IBM Research Europe - Dublin, IBM Technology Campus, Dublin 15, Ireland}
\email{niall.robertson@ie.ibm.com}

\author{Tara Murphy}
\email{tara.murphy@ibm.com}
\affiliation{IBM Quantum, IBM Research Europe - Dublin, IBM Technology Campus, Dublin 15, Ireland}

\author{Sergiy Zhuk}
\email{sergiy.zhuk@ie.ibm.com}
\affiliation{IBM Quantum, IBM Research Europe - Dublin, IBM Technology Campus, Dublin 15, Ireland}

\author{John Goold}
\email{gooldj@tcd.ie}
\affiliation{Department of Physics, Trinity College Dublin, Dublin 2, Ireland}

\begin{abstract}
Understanding how hydrodynamic behaviour emerges from the unitary evolution of the many-particle Schr{\"o}dinger equation is a central goal of non-equilibrium statistical mechanics. In this work we implement a digital simulation of the discrete time quantum dynamics of a spin-$\frac{1}{2}$ XXZ spin chain on a noisy near-term quantum device, and we extract the high temperature transport exponent at the isotropic point. We simulate the temporal decay of the relevant spin correlation function at high temperature using a pseudo-random state generated by a random circuit that is specifically tailored to the {\it ibmq-montreal} $27$ qubit device. The resulting output is a spin excitation on a highly inhomogeneous background. From the subsequent discrete time dynamics on the device we are able to extract an anomalous super-diffusive exponent consistent with the conjectured Kardar-Parisi-Zhang (KPZ) scaling at the isotropic point. Furthermore we simulate the restoration of spin diffusion with the application of an integrability breaking potential.   
\end{abstract}

\maketitle

%
%
%
%

 {\bf Introduction-} The idea that quantum dynamics of many-body physics is better simulated by controllable quantum systems was put forward by Richard Feynman 40 years ago~\cite{feynman1982simulating}. This is known as quantum simulation~\cite{lloyd1997,nielsen2002quantum} and is expected to be one of the most promising short term goals of near term quantum computing devices~\cite{Preskill2018quantumcomputingin} with inevitable applications in diverse areas ranging from quantum chemistry~\cite{kassal2011simulating,hastings2015improving,cao2019quantum} and material science~\cite{de2021materials} to high energy physics~\cite{nachman2021quantum}. Quantum simulators currently come in two different flavours: analogue and digital~\cite{Georgescu_14,Gerace_20,daley2022practical}. In an analogue simulator a purpose built controllable quantum many-body system is prepared in the laboratory with the ability to mimic a specific model Hamiltonian of interest. In a digital simulator the quantum dynamics is mapped to a series of discrete time gates that are used to directly manipulate the information encoded in the quantum state~\cite{lloyd1997}.

 While analogue simulators are built with a specific model in mind, digital simulation offers the possibility to program different Hamiltonian models so that a wide range of quantum dynamics is, in principle, accessible on the same device. The possibility of universal simulation of many-body quantum dynamics afforded by digital quantum simulation is a tantalising one. In reality, however, the current devices are still some distance from this goal with noisy gate operations and readout. Ultimately,  significant progress in error correcting techniques is needed~\cite{Preskill2018quantumcomputingin}. In fact it has been on analogue devices where the most significant progress has been made in simulating many-body dynamics~\cite{daley2022practical}. However, recent progress in error mitigation techniques for digital devices has brought us closer to getting quantitative results from noisy simulations~\cite{temme2017error,endo, kim2021scalable}. 

One dimensional interacting quantum spin systems are perhaps the simplest non-trivial models used in the field of many-body physics. Despite the obvious shortcomings on noisy near-term quantum devices, there have been several interesting digital simulations ~\cite{zhukov2018algorithmic,CerveraLierta2018exactisingmodel,francis2020quantum,smith2019simulating,vovrosh2021confinement} which are restricted to either small systems or short times. These simulations can be viewed as important benchmarks of device capability. In this work we show how noisy near-term quantum devices can be used to shed important light on a research topic which is at the forefront of research in low-dimensional quantum spin dynamics. The issue we address concerns the nature of the emergent high temperature anomalous hydrodynamics of the spin-$\frac{1}{2}$ XXZ spin chain at the isotropic point~\cite{bulchandani2021superdiffusion}.

 How macroscopic hydrodynamic behaviour emerges from  underlying microscopic physics is a question that has been at the forefront of physics for 200 years~\cite{fourier1822theorie,buchanan2005heated}. This research continues today in quantum many-body dynamics where the finite temperature transport properties of quantum spin systems is under significant analytical and numerical scrutiny~\cite{bertini_review}. A recent development was the discovery of high-temperature spin super-diffusion at the isotropic point of the spin-$\frac{1}{2}$ XXZ model~\cite{Marko_boundary_driving_XXZ} using an open systems approach. In this work the non-equilibrium steady state was found to have a current scaling $\langle\hat{J}\rangle\propto 1/\sqrt{L}$ consistent with a space time scaling $x\propto t^{1/\nu}$ with $\nu=3/2$. A numerical study of the infinite temperature spin auto correlation functions at the isotropic point~\cite{PhysRevLett.122.210602} has lead to the conjecture that the dynamics is in the KPZ universality class~\cite{kpz_original} and further numerical work~\cite{PhysRevLett.127.107201} has shown the survival of the associated anomalous scaling of the spin-spin auto-correlation functions at finite temperatures. There is still no clear consensus on the exact conditions for the emergence of this universal behaviour. Integrability is conjectured to be central in the emergence of this scaling and progress in incorporating anomalous diffusion in the context of generalised hydrodynamics~\cite{PhysRevX.11.031023,bulchandani2021superdiffusion,sarang} has been made. The predicted super-diffusive exponent has been observed in a recent experimental study of neutron scattering off $KCuF_{3}$ which realises an almost ideal XXZ spin chain~\cite{scheie2021detection}. Furthermore, the scaling was recently confirmed in two analogue simulations of spin chains in both ultra-cold atoms~\cite{wei2022quantum} and ion-trap platforms~\cite{joshi2022observing} and in polariton condensates~\cite{polariton}. 
 
 In this work we perform the first digital quantum simulation, of the discrete time dynamics, at the isotropic point of the XXZ model. It was recently discovered that the Trotterised version of the XXZ model is also integrable~\cite{PhysRevLett.121.030606} and the KPZ scaling at the isotropic point remains~\cite{PhysRevLett.122.210602,krajnik2020kardar}. This has the distinct advantage on a near term device of being able to simulate for longer times without having to worry about Trotter errors that plague continuous time simulation. We extract the high temperature correlation function following a recent proposal by Richter and Pal~\cite{PhysRevLett.126.230501,lunt2021quantum} that suggests using specially tailored pseudo-random states which are generated from a relatively shallow-depth circuit~\cite{jin2021random,arute2019quantum}. The discrete time dynamics of the spin auto-correlation function is then simulated. We apply a zero noise error mitigation strategy and remarkably show that the KPZ anomalous exponent can be extracted for over two decades of time evolution. Furthermore we show that the scaling is independent of the time period of the Trotter step and observe the restoration of diffusion, signalled by the emergence of the exponent $\nu=1/2$, when an integrability breaking staggered field perturbation is applied. 
 
{\bf Initial state preparation}
 \begin{figure}[t]
\begin{center}
\includegraphics[width=\columnwidth]{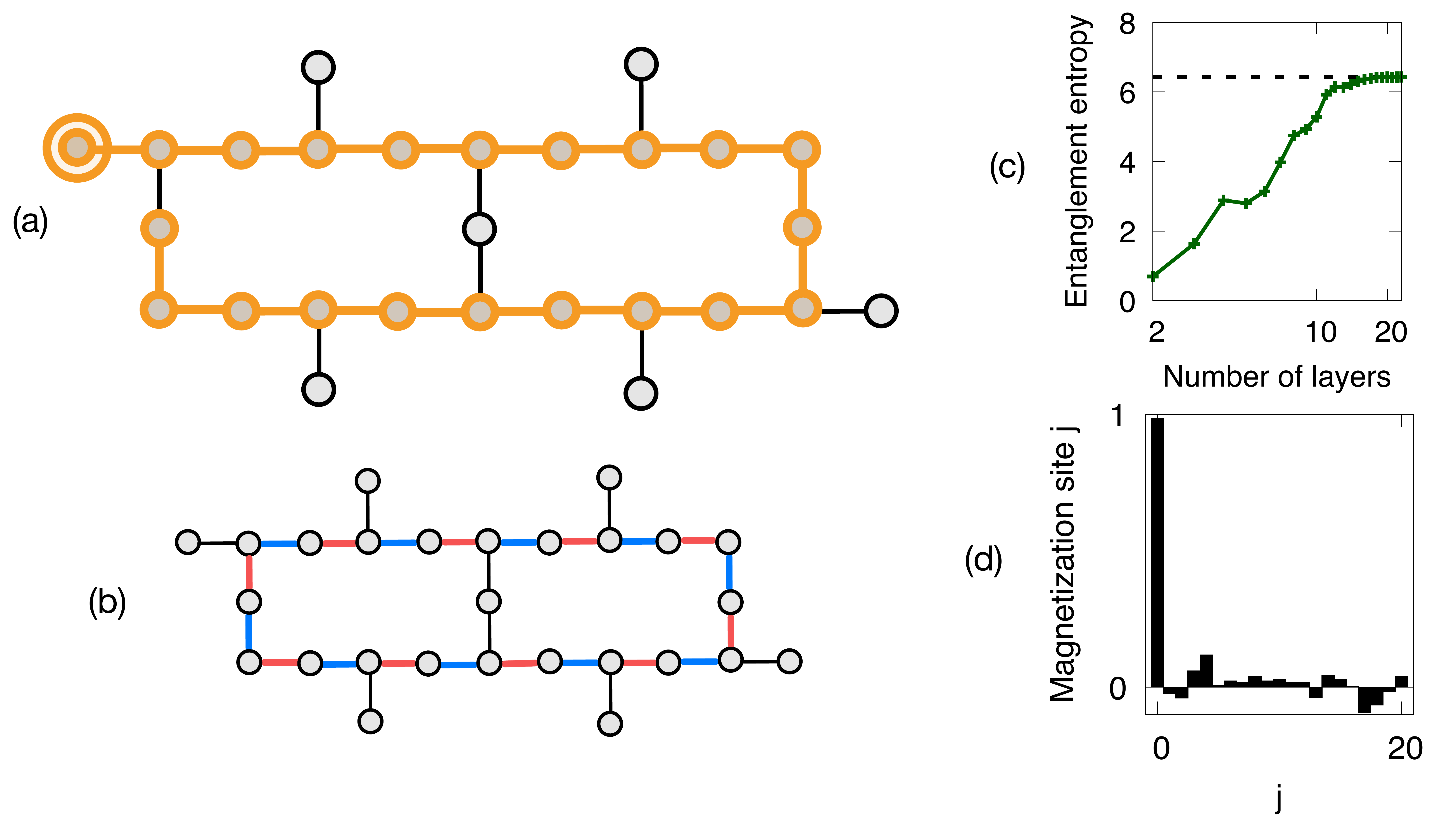}
\caption{(a) The ibmq montreal qubit connectivity, with a 1-dimensional $XXZ$ model (OBC) mapped onto a 21-qubit chain in the device (we label them $q_{j}$). Site $q_0$ is mapped to the encircled qubit, and is untouched by the randomisation procedure. (b) Red (blue) is CNOT pattern A (B) used in the random state preparation. These are alternated at each layer of the iterated random circuit. (c) The bipartite von Neumann entanglement entropy of the 20 qubit chain as a function of the number of layers in the random circuit. These results are from a clean simulation with connectivity matching that of ibmq montreal. The dashed line represents the maximum Page value~\cite{page_93}. (d) The spin density profile of the final state of one sampling of the random circuit.}
\end{center}
\end{figure}
 - All the quantum simulations in this paper were performed on the {\it ibmq montreal} $27$ qubit device based on coupled transmons. This machine was recently benchmarked to have a quantum volume of 128~\cite{ibm}. The connectivity of the device is shown in Fig 1.~(a) and we will use the 21 qubits which are shown in orange for our dynamical simulations. Our first task, following the suggestion of Richter and Pal~\cite{PhysRevLett.126.230501} is to generate a pseudo-random state state on the device leaving all but one qubit untouched ($q_0$). 
 
The randomisation procedure is split up into two sub-routines; the single qubit gate routine, and the entangling routine. A layer of the procedure is made up of a single qubit step, followed by an entangling step. The single qubit gate routine is as follows:
\begin{enumerate}
  \item At layer 1, for each qubit $q_j$, apply $G_1^j$, which is chosen randomly from the set of gates $\{X^{1/2}, Y^{1/2}, T\}$
  \item At layer $n > 1$, for each qubit $q_j$ apply $G_n^j$, which is chosen randomly from the set $\{X^{1/2}, Y^{1/2}, T\}\backslash G_{n-1}^j$
\end{enumerate}
Between each single qubit step, we carry out an entangling step. This consists of applying one of two patterns of $CX$ gates across the device. The choice of pattern in alternated between patterns ‘A’ and ‘B’ (shown in Fig 1.~(b)) at each step. The randomisation procedure is performed over multiple layers until the state is deemed sufficiently random. The number of layers that are needed is estimated from a classical simulation of the time evolution of the bi-partite entanglement of the random circuit. The results of the classical simulation are shown in Fig 1.~(c), where we show the half chain von Neumann entropy as a function of the number of layers in our preperation step. We see that already a modest number of layers is enough to saturate the maximum Page value~\cite{page_93}. Fig 1.~(d) shows the spin density profile of the final state, on the actual hardware following one sampling of the random circuit. The data was extracted by performing $30,000$ shots after one sampling of the circuit. 

In this work we will be interested in performing dynamical quantum simulation of  spin spin autocorrelation functions, which take the form
\begin{equation}
C_{jk}(t) = \text{Tr}\left(S^z_j S^z_k(t)\right)/2^L
\end{equation}
where the trace is over the entire Hilbert space. Following the proposal of Richter and Pal~\cite{PhysRevLett.126.230501,lunt2021quantum} we will use the output of our state preparation circuit in the evaluation of this object. Let us assume for a moment that the output state of the entire register would be $|\psi_{R,0}\rangle=|0\rangle|\psi_{R}\rangle$ with $|\psi_R\rangle = \sum_{n} c_n |n\rangle$, where the expansion is over the entire computational basis. If $c_n$ are Gaussian random numbers with zero mean (i.e the state is drawn randomly from the unitarity invariant Haar measure) then one can approximate the correlation functions by (for details \cite{PhysRevLett.126.230501}) 
\begin{equation}\label{corr_func_single_site}
C_{jk}(t) = \frac{1}{2}\langle\psi_{R,k}|S^z_j(t)|\psi_{R,k}\rangle+\mathcal{O} (2^{-\frac{L}{2}}).
\end{equation}
This typicality approach is routinely used to evaluate the time evolution of observables in classical simulations~\cite{PhysRevLett.112.120601,PhysRevB.99.144422,bertini_review,richter2020quantum,PhysRevB.103.184205, jin2021random}. Pseudo-random states can be now generated on noisy near-term quantum devices with relatively shallow circuits~\cite{arute2019quantum}. The state preparation  procedure leads to deviations from a Haar random state. However, as argued in \cite{PhysRevLett.126.230501,jin2021random} the exact distribution of the coefficients of the states can deviate from Gaussian and still the same result holds~\cite{jin2021random}. A key finding of \cite{PhysRevLett.126.230501} is that the state which is output after the initial randomisation phase is robust to modelled device noise. The main purpose of this paper is to use this protocol in order to extract the decay of the spin auto-correlation function on a current quantum hardware.

 {\bf Discrete time dynamics}- The spin-$\frac{1}{2}$ XXZ Hamiltonian which will be the central focus of our simulation is
\begin{equation}
    H_{XXZ} = J\sum_{\ell=0}^{L-1}\left( S^x_\ell S^x_{\ell+1} + S^y_\ell S^y_{\ell+1} + \Delta S^z_\ell S^z_{\ell+1}  \right),
\end{equation}
where $S_\ell^\alpha = \sigma^\alpha_\ell/2$ is the spin operator acting on site $\ell$ and $L$ is the number of sites. We use open boundary conditions, and focus on the isotropic point ($\Delta = 1$). We define $h_{\ell, \ell+1} = J\left(S^x_\ell S^x_{\ell+1} + S^y_\ell S^y_{\ell+1} + \Delta S^z_\ell S^z_{\ell+1}\right)$ and group all of these two-site operators into two sums: $H_1 = \sum_{\ell\text{ odd}} h_{\ell, \ell+1}$, $H_2 = \sum_{\ell\text{ even}} h_{\ell, \ell+1}$. Note that the two-site operators in a given sum all act on disjoint pairs of sites. Therefore all operators commute with all other operators in their respective sums. We now look at the discrete time dynamics given by a Trotter step $\tau$:

\begin{align}
    \mathcal{U}(n\tau) &= \left[\mathcal{U}_\text{odd}(\tau)\mathcal{U}_\text{even}(\tau)\right]^n \nonumber \\ 
    &= \left[\prod_{j=0}^{L/2-1}U_{2j, 2j+1}(\tau)\prod_{k=1}^{L/2-1}U_{2k-1, 2k}(\tau)\right]^n
\end{align}
where $U_{j k}(\tau) = e^{-ih_{j k}\tau}$. The implementation of $U_{j k}(\tau)$ in a quantum circuit is given by Fig. 2 \cite{williams}. 
\begin{figure}
    \centering
    \scalebox{0.75}{\input{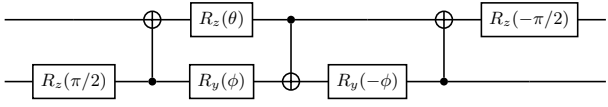}}
    \caption{Implementation of $U_{j k}(\tau)$ on $q_j$ and $q_k$. Here, $\theta = \Delta\tau/2 -\pi/2$, and $\phi = \tau/2 -\pi/2$. This circuit equals $U_{jk}(\tau)$ up to a global phase factor of $\exp(-i\pi/4)$}
    \label{fig:my_label}
\end{figure}

Note that if we keep $n\tau$ fixed and take the limit $\tau \rightarrow 0$, we get that $\mathcal{U}(n\tau) \rightarrow e^{-iH_{\text{XXZ}}n\tau}$. However, we are less interested in this trotterized unitary as an approximation of the continuous time unitary for the XXZ model, but instead as a floquet system with kicking period $\tau$. This model has Hamiltonian given by:

\begin{equation}
    H^{(\tau)}_\text{XXZ}(t) = H_1 + \tau H_2\sum_{n\in\mathbb{Z}} \delta(t-n\tau).
\end{equation}
This Hamiltonian has been shown to also give rise to KPZ like scaling in discrete time~\cite{PhysRevLett.122.210602,krajnik2020kardar} and is particularly appealing due to the fact that there is no Trotter error. This was recently exploited in a digital simulation of the spin-$\frac{1}{2}$ XXZ chain in the gapped ($\Delta>1$) phase on the {\it ibm kawasaki} $27$-qubit machine in order to study the effect of noise on conserved charges~\cite{trotter_quantum}. In our simulations we will also be interested in explicitly breaking the integrability of this model by the application of a staggered field which, at high temperatures, is expected to restore diffusion at the isotropic point. To implement this integrability breaking term, we continue like in the previous case, except as well as $H_1$ and $H_2$ we add the term $H_3 = \frac{J}{2}\sum_{\ell = 0}^{L-1}\left(-1\right)^\ell S^z_\ell$. The unitary for the discrete time evolution with the staggered field is given by:
\begin{equation}
    \mathcal{U}(n\tau) = \left[\mathcal{U}_\text{even}(\tau) \mathcal{U}_3(\tau)\mathcal{U}_\text{odd}(\tau)\right]^n,
\end{equation}
where $\mathcal{U}_3(\tau) = \prod_j e^{-i\sigma^z_j\theta_j}$ is implemented as a collection of single qubit rotations.
The effective Hamiltonian is now given by
\begin{equation}
    H^{(\tau)}_\text{stagg}(t) = H^{(\tau)}_\text{XXZ}(t) + \lim_{\epsilon\rightarrow 0^+}\tau H_3\sum_{n\in\mathbb{Z}} \delta(t-n\tau+\epsilon),
\end{equation}
where $\epsilon$ is a dummy variable used to ensure we apply the resulting unitaries in the correct order.
\begin{figure}[t]
\begin{center}
    \label{fig:my_label}
\includegraphics[width=\columnwidth]{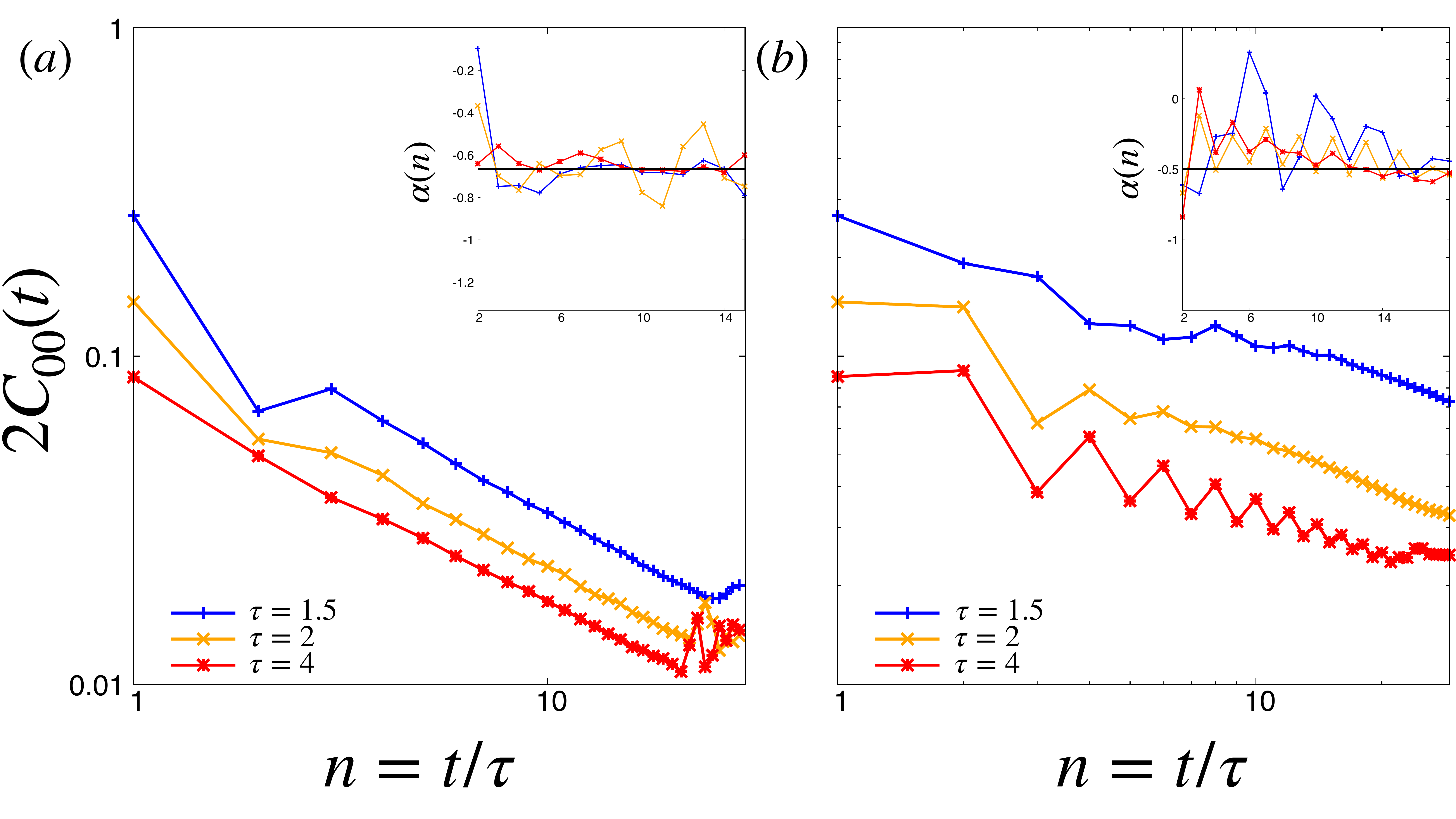}
\caption{(a) Classical trotter simulations of the spin autocorrelation function on site $0$ in the discrete time model for various different timesteps. The inset shows $\alpha(n) = \frac{d\ln C(t)}{d\ln n}$ for the same timesteps, with the black line indicating the expected super-diffusive value $-2/3$. The measured value of $\alpha$ converges towards the expected value of $-2/3$.
(b) Classical trotter simulations for the correlator in the discrete time isotropic XXZ model with a staggered field for various time steps. The inset shows $\alpha(n)$ for the same timesteps, with the black line indicating the expected diffusive exponent of $-1/2$. The measured value of $\alpha$ converges towards $-1/2$.}
\end{center}
\end{figure}

\begin{figure}
\begin{center}

\includegraphics[width=\columnwidth]{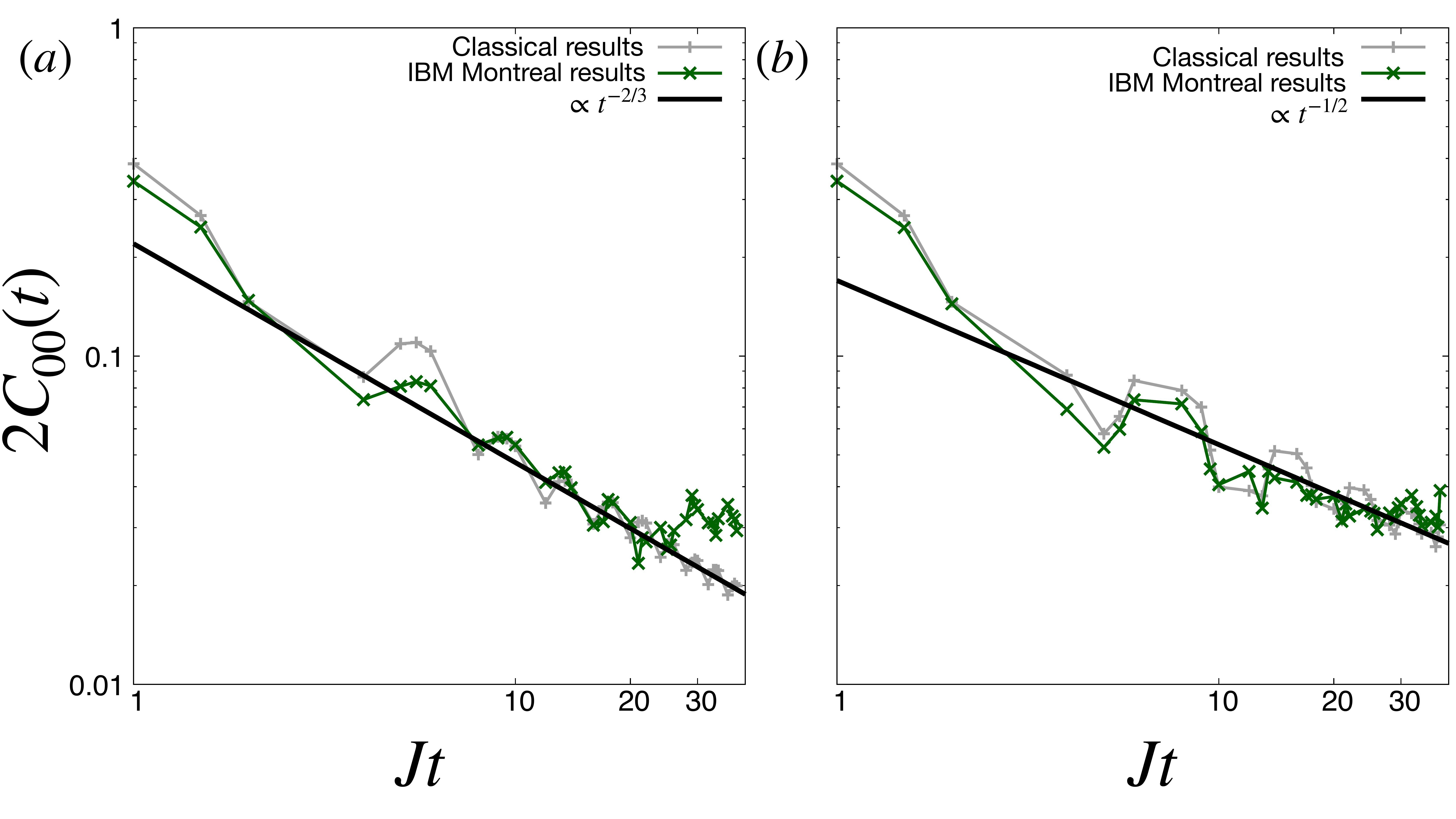}
\caption{(a) The spin autocorrelation function on site $0$ in the discrete time XXZ model at the isotropic point. The trotter step is taken to be $4J^{-1}$, with added weavings from $1J^{-1}$, $1.5J^{-1}$, and $2J^{-1}$.
(b) Results for the correlator in the discrete time isotropic XXZ model with a staggered field. The trotter step is taken to be $4J^{-1}$, with added weavings from $1J^{-1}$, $1.5J^{-1}$, and $2J^{-1}$. }
\end{center}
\end{figure}

 {\bf Results} We first demonstrate, using classical simulations~\cite{sm}, that the transport exponents at the isotropic point for both the (a) clean and (b) staggered field discrete time models are independent of step size. In Fig.~3, we do a first order trotter decomposition of the clean and staggered field models with various timesteps. The power law scaling, in both models, is found to be independent of the time step for the steps chosen. In the insets we show the oscillations of the exponent $\alpha = \frac{d\ln C(t)}{d \ln n}$ around the expected values ($2/3$ for the clean model and $1/2$ for the model with staggered fields) for each model in the insets, where $n$ is the number of Trotter steps. We avoid timesteps near $\pi$ as the transport behaviour changes drastically due to many-body resonances \cite{sm}.
 
We now come to the key finding of our work: the digital simulation on a real near term quantum computer. In Fig. 4 we show  our results for the spin auto-correlation function simulated on \textit{ibmq-montreal}. We have found that the optimal time-step for our simulations is $\tau = 4J^{-1}$~\cite{sm}. In (a), we simulate the clean model, while in (b) we add the staggered field. The green lines show the results on the quantum simulator using a first order Trotter decomposition. Remarkably our results in both the integrable and non-integrable case track the classical simulation well up to two decades in time evolution. This timescale is sufficient to see the emergence of hydrodynamic scaling. This is the main result of our work. 

In order to increase the number of data points for our power law fit we have employed the concept of weaving~\cite{weave}. This allows us to look at more data points in time. The idea is to  artificially increase our time resolution in our study of the the floquet unitary $\mathcal{U}(n\tau)$~\cite{sm}. This is done by adding smaller $\mathcal{U}(\tau' < \tau)$ at the start of the circuit as a modified initial condition, and then weaving the evolution of this new initial state (shifted slightly in time) with the original evolution. We add weaves of $1J^{-1}, 1.5J^{-1}, 2J^{-1}$. Furthermore, in obtaining these results, we employ a form of error mitigation known as `zero noise extrapolation', or \textit{zne}~\cite{temme2017error}. However, we do not find that it significantly helps at these time scales for our first order Trotter simulation~\cite{sm}. 

To extract the power law behaviour of the results from the quantum simulations, we analyse the intersection of two regimes in time: 1) Where the power law scaling is present in the classical results, and 2) where the quantum results have little error compared to the classical results. We then fit a power law to the quantum results via least squares. For panel (a) with the clean  model, we get that $\alpha \approx -0.644$, and for panel (b) with the staggered external field we get that $\alpha \approx -0.505$. These have relative errors $\sim 3.40\%$ and $\sim 1.00\%$ compared to the expected scalings of $-2/3$ and $-1/2$ respectively. 

IBM's quantum devices are calibrated regularly. When these results were collected, the readout error of qubit $0$ was $2.26\%$, while the average error of all the CNOTs (excluding an outlier at $18.4\%$) used in the simulation was $1.12\%$, with a standard deviation of $0.52\%$. For more details see~\cite{ibm,sm}.

{\bf Discussion---} We have provided strong evidence that KPZ scaling and the restoration of diffusion through explicit integrability breaking can  be simulated digitally on a near term device. Our work is inspired by the proposal of Richter and Pal~\cite{PhysRevLett.126.230501} which exploits a pseudo-random state as a starting point for the simulation. It is remarkable that our digital quantum simulation is able to follow closely the classical simulation to over two decades in time evolution. There are several features of this simulation which are worth pointing out. First of all the nature of the initial state appears to be extremely useful for the extraction of infinite temperature transport exponents on current quantum hardware. The precise interplay between noise channels and such pseudo-typical states merits future detailed investigations. Since these states are locally equivalent to the identity, it is plausible that they offer a special resilience to unital channels such as dephasing. We have confirmed, on hardware, the suggestion~\cite{PhysRevLett.126.230501} that the hydrodynamic scaling is accessible despite inevitable device noise. Secondly and most importantly the key feature of our simulations is that we work with the discrete time model and this allows us to simulate long times without Trotter error~\cite{PhysRevLett.122.210602,PhysRevLett.121.030606}.  To our knowledge this is the first extraction of transport exponents of an interacting quantum system on a digital quantum device. Our findings are consistent with recent experiments in a variety of physical platforms~\cite{scheie2021detection,wei2022quantum,joshi2022observing,polariton}. As hardware improves further and the number of good device qubits increase, we hope that our work will inspire further work on high temperature transport of non-integrable and integrable  quantum many-body models in regimes not accessible to classical numerics.

{\bf Acknowledgements.---} We thank the QuSys group at TCD, A.~Purkayastha and A. Silva for useful discussions. JG is funded by  a Science Foundation Ireland-Royal Society University Research Fellowship, the European Research Council Starting Grant ODYSSEY (Grant Agreement No. 758403). This project was made possible through the newly established TCD-IBM predoctoral programme. 

\section{Appendix}

\setcounter{figure}{0}
\renewcommand{\thefigure}{A\arabic{figure}}

\setcounter{equation}{0}
\renewcommand{\theequation}{A\arabic{equation}}

\subsection{A1. Stochastic Trace Evaluation}
The random circuit preparation we use we creates a pseudo Haar state state $|\psi_R\rangle$ which is known in the literature to provide a stochastic evaluation of the trace such that ~\cite{jin2021random,arute2019quantum, lunt2021quantum, PhysRevLett.126.230501}
\begin{equation}
    \mathbb{E}(\langle |\psi_R|O | |\psi_R\rangle) = \text{Tr}(O)/D + \mathcal{O}(1/\sqrt{D}),
\end{equation}
for some observable $O$, where $D$ is the dimension of the Hilbert space.
Furthermore, \cite{PhysRevLett.126.230501} shows that the stochastic trace evaluation is resilient to typical noise processes on hardware. 
The circuit we use in the main text prepares the initial state $|\psi_{R, 0}\rangle$. Qubit $0$ remains idle while the other $20$ qubits are randomised via the procedure outlined in the main text. The state then goes through the simulation part of the circuit, followed by a $Z$ measurement on qubit $0$. So the resulting object we calculate is:

\begin{align}
    \mathbb{E}\left(\langle\psi_{R, 0}| S^z_0(t)| \psi_{R, 0}\rangle\right) &= 2\mathbb{E}\left(\langle\psi_R| P^0_0 S^z_0(t) P^0_0 |\psi_R \rangle\right) \nonumber \\
    &= \text{Tr}\left(P^0_0 S^z_0(t) P^0_0\right)/2^{L-1} \nonumber
\end{align}
where $|\psi_R\rangle$ is a state fully randomised on all qubits, and $P_0^0$ is the $|0\rangle$ projector on qubit $0$. Note that $||P^0_0|\psi_R\rangle||^2 \approx 1/2$, so we pick up a factor of $2$.
We now use the cyclic property of the trace, with the fact that $(P_0^0)^2 = P_0^0$ to show that
\begin{align}
    &\mathbb{E}\left(\langle\psi_{R, 0}| S^z_0(t)| \psi_{R, 0}\rangle\right)
    \nonumber \\
    &= \text{Tr}\left((P^0_0)^2 S^z_0(t)\right)/2^{L-1} = \text{Tr}\left(P^0_0 S^z_0(t)\right)/2^{L-1} \nonumber \\
    &= \text{Tr}\left(P^0_0 S^z_0(t) - S^z_0(t)/2\right)/2^{L-1} \nonumber \\
    &= \text{Tr}\left((P^0_0 - 1/2) S^z_0(t)\right)/2^{L-1} = \text{Tr}\left(S^z_0S^z_0(t)\right)/2^{L-1}. \nonumber \\
    &= 2C_{00}(t),
\end{align}
where $C_{jk}(t) = \text{Tr}\left(S^z_j S^z_k(t)\right)/2^L$. This demonstrates that such a state can be used to approximate high temperature spin spin correlation functions.

\subsection{A2. Error Mitigation}
In an attempt to increase the accuracy of our simulation on  {\it ibmq-montreal} we use error mitigation. In particular, we use a form of zero-noise extrapolation (ZNE) \cite{temme2017error, li2017efficient}. The idea behind ZNE is quite intuitive; if one assumes that the expectation value of interest is an analytical function of some noise parameter $\lambda$, then running the circuit for a number of different values of $\lambda$ allows us to use some extrapolation technique (e.g. Richardson, linear, exponential) to obtain an estimate for the expectation value $E(\lambda)$ with zero noise. Here, $\lambda$ is a parameter that characterises the level of noise in the circuit. As such, $E(\lambda)$ is the evaluation of the expectation value in question with noise level $\lambda$. There are a number of different ways to artificially increase the noise of the circuit. If one assumes that the noise is constant in time, then the ideal procedure is to stretch the control pulses in time \cite{kim2021scalable}. Alternatively, one can increase the runtime at a higher level of abstraction, i.e. circuit design, by adding layers of gates that are equivalent to the identity \cite{giurgica2020digital}. An example of such an approach is \textit{unitary folding}, where if the circuit is written as a unitary $U$, one applies $U\left[U^{\dagger}U\right]^f$ for a number of different integers $f$ to increase the depth and hence the runtime of the circuit. In this work, we use \textit{partial unitary folding} which is particularly suited to circuits based on Trotterisation. The key idea here is to apply the folding technique only to a fixed number of Trotter steps instead of the entire unitary $U$. 

We focus on exponential extrapolation \cite{giurgica2020digital}, in which case $E(\lambda)$ is modelled in the following way:
\begin{equation}
    E(\lambda) = a + b e^{-c\lambda}.
\end{equation}
Technically in the calculation of $E(\lambda)$ we need to take into consideration finite sampling errors due to taking a finite number of shots. In our case, we take enough shots to ignore this compared to the errors from circuit noise.

One can further simplify the above noise model by arguing that $a = 0$ from the observation that with infinite noise, the quantum state $\rho$ would become equivalent to the identity and thus $E(\lambda = \infty) = \text{Tr}\left(\sigma^z_0\right) = 0$.  

{\bf Unitary Folding} - We artificially increase the noise of our circuit via unitary-folding, and partial unitary-folding. The idea of unitary-folding is to increase the noise of the circuit by using the same gates present in the circuit, along with their daggers in order to increase the depth of the circuit, without changing the effective unitary of the circuit. We do not include the state preparation part of the circuit in this process, as the randomisation procedure that we employ that was suggested in~\cite{PhysRevLett.126.230501} is robust against noise for the purpose of stochastic trace evaluation. Let the effective unitary of the simulation part of the circuit be given by $U$. We can then fold $U$ on top of itself by implementing the following simulation unitary:
\begin{equation}
    U\left[U^\dagger U\right]^f,
\end{equation}
which is $2f+1$ times deeper than the original simulation circuit. The noise factor of this circuit is thus $2f+1$. So we can now evaluate $E(\lambda)$ at $\lambda = 3, 5, 7,...$. The problem is that noise quickly dominates for deeper circuits, so sampling bigger and bigger values of $\lambda$ doesn't give our regression model much info about $E(0)$. Instead, we can have a finer graining in $lambda$ if we instead only {\it partially} fold the unitary. This method is extremely suited to Trotter evolution, as we can break the circuit into Trotter layers (all of which use the same device gates), and only fold a subset of these layers instead of the entire circuit. Let $0 \leq s \leq m$ where there are a total of $m$ Trotter layers in the circuit. We can then implement the circuit:
\begin{equation}
    U\left[U^\dagger U\right]^f T_m^\dagger T_{m-1}^\dagger ... T_{m-s+1}^\dagger T_{m-s+1} ... T_{m-1} T_m, 
\end{equation}
which has noise factor $\lambda = 2f + 1 + \frac{2s}{m}$, giving us a much finer graining than just full unitary-folding alone. All of our simulation circuits for different time steps are folded for an array of values between $\lambda = 1$ and $\lambda = 3$.
\begin{figure}[t]
\begin{center}
\label{fig:zne}
\includegraphics[width=\columnwidth]{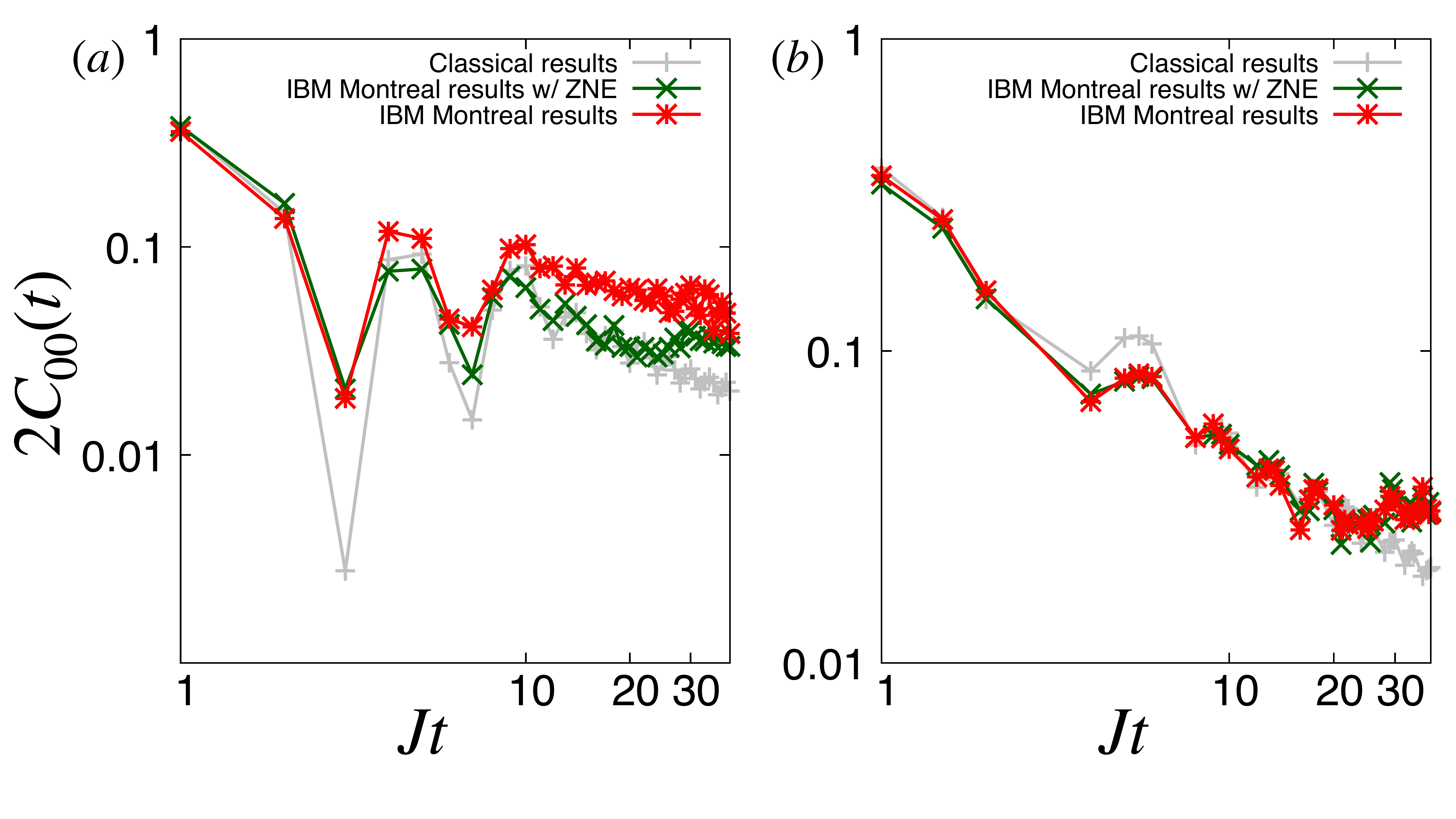}
\caption{(a) Original simulations with 2nd order Trotter with $J\tau = 4$ and weavings from $J\tau = 1, 2, 3$. Our ZNE protocol clearly improves our results here.
(b) Final simulations with 1st order Trotter with $J\tau = 4$ and weavings from $J\tau = 1, 1.5, 2$. There is no discernible advantage of using ZNE in these simulations.}
\end{center}
\end{figure}

The ZNE technique seems to work well to correct noise from higher order Trotter expansions used for simulating continuous time dynamics. For example for 2nd order Trotter shown in Fig. S1 (a) we demonstrate that it provides a worth-while exploitable trade off for accuracy vs circuit depth compared to 1st order (depth scales linearly in number of Trotter steps with same leading coefficient in both cases). The aim of our paper was to study KPZ scaling which preserved in the discrete time dynamics \cite{PhysRevLett.122.210602,krajnik2020kardar}. In the case studied in the main text we found that the ZNE procedure did not provide any noticeable advantage.
\begin{figure}
\begin{center}
    \label{fig:phase}
\includegraphics[width=\columnwidth]{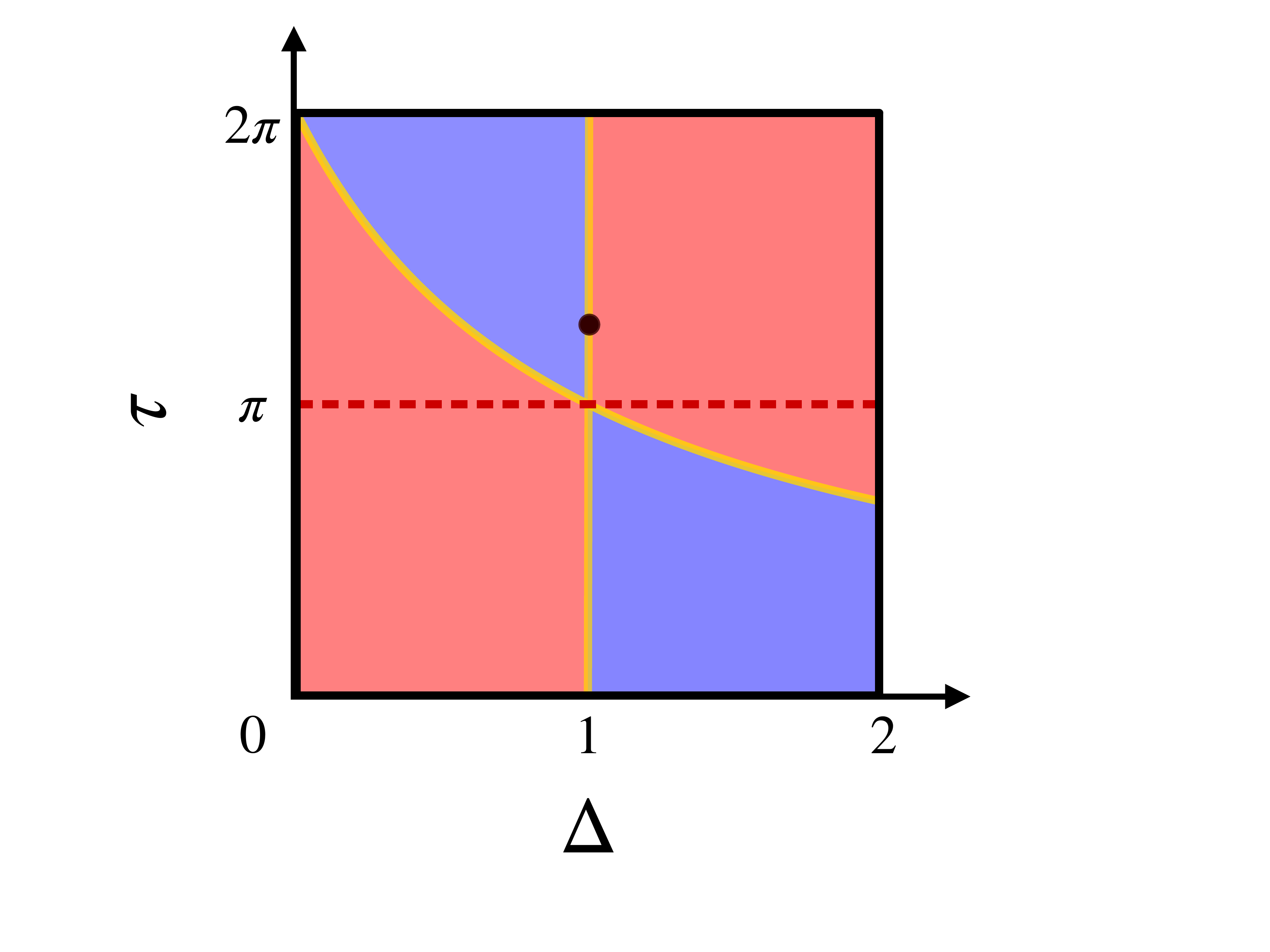}
\caption{Red: ballistic, blue: diffusive, yellow: superdiffusive, dotted red: dual unitary point (ballistic). The black dot indicates the part of the phase diagram that we were probing in our experiments.}
\end{center}
\end{figure}

One might also notice another difference in (a) and (b) of Fig. S1. We switch from using weavings $(1, 2, 3)$ to weavings $(1, 1.5, 2)$. This is due to big dips that can be seen in Fig. S1 corresponding to the $\tau_{0} = 3J^{-1}$ strand of the weave. To avoid this weird behaviour while keeping the average resolution the same, we pick unequal spacings for the weave instead, and as one can see these big dips are avoided. These dips are actually due to the simulated model being near the dual unitary point of discrete XXZ at $J\tau = \pi$ \cite{ljubotina2019ballistic}.

\subsection{A3. Other Comments}
We chose to study the isotropic point of the XXZ chain as it convenient in the sense that the super-diffusive scaling can be observed with significant Trotter step sizes. In \cite{ljubotina2019ballistic}, the authors study the transport phase diagram of the same Floquet operator as in this paper using a slightly different Hamiltonian parameterisation
\begin{equation}
    \mathcal{J}\left(\sigma^x\otimes\sigma^x + \sigma^y\otimes\sigma^y\right) + \mathcal{J}'\sigma^z\otimes\sigma^z,
\end{equation}.
The study the transport regimes as a function of $\mathcal{J}$ and $\mathcal{J}'$. Let us explicitly compare our results for the purpose of consistency. In our case, the Hamiltonian for the two qubit interaction is given by:
\begin{equation}
    \frac{J\tau}{4}\left(\sigma^x\otimes\sigma^x + \sigma^y\otimes\sigma^y + \Delta \sigma^z\otimes\sigma^z\right),
\end{equation}
so we change variables to:
\begin{align}
    \tau &= 4\mathcal{J}, \\
    \Delta &= \mathcal{J}'/\mathcal{J},
\end{align}
where $\tau$ is given in units of $J^{-1}$. The black dot in Fig. S2 shows the point of the phase diagram discussed in \cite{ljubotina2019ballistic} that we were probing in our experiments. 
\bibliography{IBM.bib}

\begin{thebibliography}{53}%
\makeatletter
\providecommand \@ifxundefined [1]{%
 \@ifx{#1\undefined}
}%
\providecommand \@ifnum [1]{%
 \ifnum #1\expandafter \@firstoftwo
 \else \expandafter \@secondoftwo
 \fi
}%
\providecommand \@ifx [1]{%
 \ifx #1\expandafter \@firstoftwo
 \else \expandafter \@secondoftwo
 \fi
}%
\providecommand \natexlab [1]{#1}%
\providecommand \enquote  [1]{``#1''}%
\providecommand \bibnamefont  [1]{#1}%
\providecommand \bibfnamefont [1]{#1}%
\providecommand \citenamefont [1]{#1}%
\providecommand \href@noop [0]{\@secondoftwo}%
\providecommand \href [0]{\begingroup \@sanitize@url \@href}%
\providecommand \@href[1]{\@@startlink{#1}\@@href}%
\providecommand \@@href[1]{\endgroup#1\@@endlink}%
\providecommand \@sanitize@url [0]{\catcode `\\12\catcode `\$12\catcode
  `\&12\catcode `\#12\catcode `\^12\catcode `\_12\catcode `\%12\relax}%
\providecommand \@@startlink[1]{}%
\providecommand \@@endlink[0]{}%
\providecommand \url  [0]{\begingroup\@sanitize@url \@url }%
\providecommand \@url [1]{\endgroup\@href {#1}{\urlprefix }}%
\providecommand \urlprefix  [0]{URL }%
\providecommand \Eprint [0]{\href }%
\providecommand \doibase [0]{http://dx.doi.org/}%
\providecommand \selectlanguage [0]{\@gobble}%
\providecommand \bibinfo  [0]{\@secondoftwo}%
\providecommand \bibfield  [0]{\@secondoftwo}%
\providecommand \translation [1]{[#1]}%
\providecommand \BibitemOpen [0]{}%
\providecommand \bibitemStop [0]{}%
\providecommand \bibitemNoStop [0]{.\EOS\space}%
\providecommand \EOS [0]{\spacefactor3000\relax}%
\providecommand \BibitemShut  [1]{\csname bibitem#1\endcsname}%
\let\auto@bib@innerbib\@empty
\bibitem [{\citenamefont {Feynman}(1982)}]{feynman1982simulating}%
  \BibitemOpen
  \bibfield  {author} {\bibinfo {author} {\bibfnamefont {R.~P.}\ \bibnamefont
  {Feynman}},\ }\href {https://doi.org/10.1007/BF02650179} {\bibfield
  {journal} {\bibinfo  {journal} {International Journal of Theoretical
  Physics}\ }\textbf {\bibinfo {volume} {21}} (\bibinfo {year}
  {1982})}\BibitemShut {NoStop}%
\bibitem [{\citenamefont {Lloyd}(1996)}]{lloyd1997}%
  \BibitemOpen
  \bibfield  {author} {\bibinfo {author} {\bibfnamefont {S.}~\bibnamefont
  {Lloyd}},\ }\href {\doibase 10.1126/science.273.5278.1073} {\bibfield
  {journal} {\bibinfo  {journal} {Science}\ }\textbf {\bibinfo {volume}
  {273}},\ \bibinfo {pages} {1073} (\bibinfo {year} {1996})}\BibitemShut
  {NoStop}%
\bibitem [{\citenamefont {Nielsen}\ and\ \citenamefont
  {Chuang}(2002)}]{nielsen2002quantum}%
  \BibitemOpen
  \bibfield  {author} {\bibinfo {author} {\bibfnamefont {M.~A.}\ \bibnamefont
  {Nielsen}}\ and\ \bibinfo {author} {\bibfnamefont {I.}~\bibnamefont
  {Chuang}},\ }\href
  {https://www.cambridge.org/highereducation/books/quantum-computation-and-quantum-information/01E10196D0A682A6AEFFEA52D53BE9AE}
  {\enquote {\bibinfo {title} {Quantum computation and quantum information},}\
  } (\bibinfo {year} {2002})\BibitemShut {NoStop}%
\bibitem [{\citenamefont {Preskill}(2018)}]{Preskill2018quantumcomputingin}%
  \BibitemOpen
  \bibfield  {author} {\bibinfo {author} {\bibfnamefont {J.}~\bibnamefont
  {Preskill}},\ }\href {\doibase 10.22331/q-2018-08-06-79} {\bibfield
  {journal} {\bibinfo  {journal} {{Quantum}}\ }\textbf {\bibinfo {volume}
  {2}},\ \bibinfo {pages} {79} (\bibinfo {year} {2018})}\BibitemShut {NoStop}%
\bibitem [{\citenamefont {Kassal}\ \emph {et~al.}(2011)\citenamefont {Kassal},
  \citenamefont {Whitfield}, \citenamefont {Perdomo-Ortiz}, \citenamefont
  {Yung},\ and\ \citenamefont {Aspuru-Guzik}}]{kassal2011simulating}%
  \BibitemOpen
  \bibfield  {author} {\bibinfo {author} {\bibfnamefont {I.}~\bibnamefont
  {Kassal}}, \bibinfo {author} {\bibfnamefont {J.~D.}\ \bibnamefont
  {Whitfield}}, \bibinfo {author} {\bibfnamefont {A.}~\bibnamefont
  {Perdomo-Ortiz}}, \bibinfo {author} {\bibfnamefont {M.-H.}\ \bibnamefont
  {Yung}}, \ and\ \bibinfo {author} {\bibfnamefont {A.}~\bibnamefont
  {Aspuru-Guzik}},\ }\href
  {https://doi.org/10.1146/annurev-physchem-032210-103512} {\bibfield
  {journal} {\bibinfo  {journal} {Annual review of physical chemistry}\
  }\textbf {\bibinfo {volume} {62}},\ \bibinfo {pages} {185} (\bibinfo {year}
  {2011})}\BibitemShut {NoStop}%
\bibitem [{\citenamefont {Hastings}\ \emph {et~al.}(2015)\citenamefont
  {Hastings}, \citenamefont {Wecker}, \citenamefont {Bauer},\ and\
  \citenamefont {Troyer}}]{hastings2015improving}%
  \BibitemOpen
  \bibfield  {author} {\bibinfo {author} {\bibfnamefont {M.~B.}\ \bibnamefont
  {Hastings}}, \bibinfo {author} {\bibfnamefont {D.}~\bibnamefont {Wecker}},
  \bibinfo {author} {\bibfnamefont {B.}~\bibnamefont {Bauer}}, \ and\ \bibinfo
  {author} {\bibfnamefont {M.}~\bibnamefont {Troyer}},\ }\href
  {https://dl.acm.org/doi/10.5555/2685188.2685189} {\bibfield  {journal}
  {\bibinfo  {journal} {Quantum Information \& Computation}\ }\textbf {\bibinfo
  {volume} {15}},\ \bibinfo {pages} {1} (\bibinfo {year} {2015})}\BibitemShut
  {NoStop}%
\bibitem [{\citenamefont {Cao}\ \emph {et~al.}(2019)\citenamefont {Cao},
  \citenamefont {Romero}, \citenamefont {Olson}, \citenamefont {Degroote},
  \citenamefont {Johnson}, \citenamefont {Kieferov{\'a}}, \citenamefont
  {Kivlichan}, \citenamefont {Menke}, \citenamefont {Peropadre}, \citenamefont
  {Sawaya} \emph {et~al.}}]{cao2019quantum}%
  \BibitemOpen
  \bibfield  {author} {\bibinfo {author} {\bibfnamefont {Y.}~\bibnamefont
  {Cao}}, \bibinfo {author} {\bibfnamefont {J.}~\bibnamefont {Romero}},
  \bibinfo {author} {\bibfnamefont {J.~P.}\ \bibnamefont {Olson}}, \bibinfo
  {author} {\bibfnamefont {M.}~\bibnamefont {Degroote}}, \bibinfo {author}
  {\bibfnamefont {P.~D.}\ \bibnamefont {Johnson}}, \bibinfo {author}
  {\bibfnamefont {M.}~\bibnamefont {Kieferov{\'a}}}, \bibinfo {author}
  {\bibfnamefont {I.~D.}\ \bibnamefont {Kivlichan}}, \bibinfo {author}
  {\bibfnamefont {T.}~\bibnamefont {Menke}}, \bibinfo {author} {\bibfnamefont
  {B.}~\bibnamefont {Peropadre}}, \bibinfo {author} {\bibfnamefont {N.~P.}\
  \bibnamefont {Sawaya}},  \emph {et~al.},\ }\href
  {https://pubs.acs.org/doi/10.1021/acs.chemrev.8b00803} {\bibfield  {journal}
  {\bibinfo  {journal} {Chemical reviews}\ }\textbf {\bibinfo {volume} {119}},\
  \bibinfo {pages} {10856} (\bibinfo {year} {2019})}\BibitemShut {NoStop}%
\bibitem [{\citenamefont {de~Leon}\ \emph {et~al.}(2021)\citenamefont
  {de~Leon}, \citenamefont {Itoh}, \citenamefont {Kim}, \citenamefont {Mehta},
  \citenamefont {Northup}, \citenamefont {Paik}, \citenamefont {Palmer},
  \citenamefont {Samarth}, \citenamefont {Sangtawesin},\ and\ \citenamefont
  {Steuerman}}]{de2021materials}%
  \BibitemOpen
  \bibfield  {author} {\bibinfo {author} {\bibfnamefont {N.~P.}\ \bibnamefont
  {de~Leon}}, \bibinfo {author} {\bibfnamefont {K.~M.}\ \bibnamefont {Itoh}},
  \bibinfo {author} {\bibfnamefont {D.}~\bibnamefont {Kim}}, \bibinfo {author}
  {\bibfnamefont {K.~K.}\ \bibnamefont {Mehta}}, \bibinfo {author}
  {\bibfnamefont {T.~E.}\ \bibnamefont {Northup}}, \bibinfo {author}
  {\bibfnamefont {H.}~\bibnamefont {Paik}}, \bibinfo {author} {\bibfnamefont
  {B.}~\bibnamefont {Palmer}}, \bibinfo {author} {\bibfnamefont
  {N.}~\bibnamefont {Samarth}}, \bibinfo {author} {\bibfnamefont
  {S.}~\bibnamefont {Sangtawesin}}, \ and\ \bibinfo {author} {\bibfnamefont
  {D.}~\bibnamefont {Steuerman}},\ }\href
  {https://www.science.org/doi/10.1126/science.abb2823} {\bibfield  {journal}
  {\bibinfo  {journal} {Science}\ }\textbf {\bibinfo {volume} {372}},\ \bibinfo
  {pages} {eabb2823} (\bibinfo {year} {2021})}\BibitemShut {NoStop}%
\bibitem [{\citenamefont {Nachman}\ \emph {et~al.}(2021)\citenamefont
  {Nachman}, \citenamefont {Provasoli}, \citenamefont {De~Jong},\ and\
  \citenamefont {Bauer}}]{nachman2021quantum}%
  \BibitemOpen
  \bibfield  {author} {\bibinfo {author} {\bibfnamefont {B.}~\bibnamefont
  {Nachman}}, \bibinfo {author} {\bibfnamefont {D.}~\bibnamefont {Provasoli}},
  \bibinfo {author} {\bibfnamefont {W.~A.}\ \bibnamefont {De~Jong}}, \ and\
  \bibinfo {author} {\bibfnamefont {C.~W.}\ \bibnamefont {Bauer}},\ }\href
  {https://link.aps.org/doi/10.1103/PhysRevLett.126.062001} {\bibfield
  {journal} {\bibinfo  {journal} {Physical review letters}\ }\textbf {\bibinfo
  {volume} {126}},\ \bibinfo {pages} {062001} (\bibinfo {year}
  {2021})}\BibitemShut {NoStop}%
\bibitem [{\citenamefont {Georgescu}\ \emph {et~al.}(2014)\citenamefont
  {Georgescu}, \citenamefont {Ashhab},\ and\ \citenamefont
  {Nori}}]{Georgescu_14}%
  \BibitemOpen
  \bibfield  {author} {\bibinfo {author} {\bibfnamefont {I.~M.}\ \bibnamefont
  {Georgescu}}, \bibinfo {author} {\bibfnamefont {S.}~\bibnamefont {Ashhab}}, \
  and\ \bibinfo {author} {\bibfnamefont {F.}~\bibnamefont {Nori}},\ }\href
  {\doibase 10.1103/RevModPhys.86.153} {\bibfield  {journal} {\bibinfo
  {journal} {Rev. Mod. Phys.}\ }\textbf {\bibinfo {volume} {86}},\ \bibinfo
  {pages} {153} (\bibinfo {year} {2014})}\BibitemShut {NoStop}%
\bibitem [{\citenamefont {Tacchino}\ \emph {et~al.}(2020)\citenamefont
  {Tacchino}, \citenamefont {Chiesa}, \citenamefont {Carretta},\ and\
  \citenamefont {Gerace}}]{Gerace_20}%
  \BibitemOpen
  \bibfield  {author} {\bibinfo {author} {\bibfnamefont {F.}~\bibnamefont
  {Tacchino}}, \bibinfo {author} {\bibfnamefont {A.}~\bibnamefont {Chiesa}},
  \bibinfo {author} {\bibfnamefont {S.}~\bibnamefont {Carretta}}, \ and\
  \bibinfo {author} {\bibfnamefont {D.}~\bibnamefont {Gerace}},\ }\href
  {https://onlinelibrary.wiley.com/doi/abs/10.1002/qute.201900052} {\bibfield
  {journal} {\bibinfo  {journal} {Advanced Quantum Technologies}\ }\textbf
  {\bibinfo {volume} {3}},\ \bibinfo {pages} {1900052} (\bibinfo {year}
  {2020})}\BibitemShut {NoStop}%
\bibitem [{\citenamefont {Daley}\ \emph {et~al.}(2022)\citenamefont {Daley},
  \citenamefont {Bloch}, \citenamefont {Kokail}, \citenamefont {Flannigan},
  \citenamefont {Pearson}, \citenamefont {Troyer},\ and\ \citenamefont
  {Zoller}}]{daley2022practical}%
  \BibitemOpen
  \bibfield  {author} {\bibinfo {author} {\bibfnamefont {A.~J.}\ \bibnamefont
  {Daley}}, \bibinfo {author} {\bibfnamefont {I.}~\bibnamefont {Bloch}},
  \bibinfo {author} {\bibfnamefont {C.}~\bibnamefont {Kokail}}, \bibinfo
  {author} {\bibfnamefont {S.}~\bibnamefont {Flannigan}}, \bibinfo {author}
  {\bibfnamefont {N.}~\bibnamefont {Pearson}}, \bibinfo {author} {\bibfnamefont
  {M.}~\bibnamefont {Troyer}}, \ and\ \bibinfo {author} {\bibfnamefont
  {P.}~\bibnamefont {Zoller}},\ }\href
  {https://www.nature.com/articles/s41586-022-04940-6} {\bibfield  {journal}
  {\bibinfo  {journal} {Nature}\ }\textbf {\bibinfo {volume} {607}},\ \bibinfo
  {pages} {667} (\bibinfo {year} {2022})}\BibitemShut {NoStop}%
\bibitem [{\citenamefont {Temme}\ \emph {et~al.}(2017)\citenamefont {Temme},
  \citenamefont {Bravyi},\ and\ \citenamefont {Gambetta}}]{temme2017error}%
  \BibitemOpen
  \bibfield  {author} {\bibinfo {author} {\bibfnamefont {K.}~\bibnamefont
  {Temme}}, \bibinfo {author} {\bibfnamefont {S.}~\bibnamefont {Bravyi}}, \
  and\ \bibinfo {author} {\bibfnamefont {J.~M.}\ \bibnamefont {Gambetta}},\
  }\href {https://link.aps.org/doi/10.1103/PhysRevLett.119.180509} {\bibfield
  {journal} {\bibinfo  {journal} {Physical review letters}\ }\textbf {\bibinfo
  {volume} {119}},\ \bibinfo {pages} {180509} (\bibinfo {year}
  {2017})}\BibitemShut {NoStop}%
\bibitem [{\citenamefont {Endo}\ \emph {et~al.}(2018)\citenamefont {Endo},
  \citenamefont {Benjamin},\ and\ \citenamefont {Li}}]{endo}%
  \BibitemOpen
  \bibfield  {author} {\bibinfo {author} {\bibfnamefont {S.}~\bibnamefont
  {Endo}}, \bibinfo {author} {\bibfnamefont {S.~C.}\ \bibnamefont {Benjamin}},
  \ and\ \bibinfo {author} {\bibfnamefont {Y.}~\bibnamefont {Li}},\ }\href
  {\doibase 10.1103/PhysRevX.8.031027} {\bibfield  {journal} {\bibinfo
  {journal} {Phys. Rev. X}\ }\textbf {\bibinfo {volume} {8}},\ \bibinfo {pages}
  {031027} (\bibinfo {year} {2018})}\BibitemShut {NoStop}%
\bibitem [{\citenamefont {Kim}\ \emph {et~al.}(2021)\citenamefont {Kim},
  \citenamefont {Wood}, \citenamefont {Yoder}, \citenamefont {Merkel},
  \citenamefont {Gambetta}, \citenamefont {Temme},\ and\ \citenamefont
  {Kandala}}]{kim2021scalable}%
  \BibitemOpen
  \bibfield  {author} {\bibinfo {author} {\bibfnamefont {Y.}~\bibnamefont
  {Kim}}, \bibinfo {author} {\bibfnamefont {C.~J.}\ \bibnamefont {Wood}},
  \bibinfo {author} {\bibfnamefont {T.~J.}\ \bibnamefont {Yoder}}, \bibinfo
  {author} {\bibfnamefont {S.~T.}\ \bibnamefont {Merkel}}, \bibinfo {author}
  {\bibfnamefont {J.~M.}\ \bibnamefont {Gambetta}}, \bibinfo {author}
  {\bibfnamefont {K.}~\bibnamefont {Temme}}, \ and\ \bibinfo {author}
  {\bibfnamefont {A.}~\bibnamefont {Kandala}},\ }\href
  {https://arxiv.org/abs/2108.09197} {\bibfield  {journal} {\bibinfo  {journal}
  {arXiv preprint arXiv:2108.09197}\ } (\bibinfo {year} {2021})}\BibitemShut
  {NoStop}%
\bibitem [{\citenamefont {Zhukov}\ \emph {et~al.}(2018)\citenamefont {Zhukov},
  \citenamefont {Remizov}, \citenamefont {Pogosov},\ and\ \citenamefont
  {Lozovik}}]{zhukov2018algorithmic}%
  \BibitemOpen
  \bibfield  {author} {\bibinfo {author} {\bibfnamefont {A.}~\bibnamefont
  {Zhukov}}, \bibinfo {author} {\bibfnamefont {S.}~\bibnamefont {Remizov}},
  \bibinfo {author} {\bibfnamefont {W.}~\bibnamefont {Pogosov}}, \ and\
  \bibinfo {author} {\bibfnamefont {Y.~E.}\ \bibnamefont {Lozovik}},\ }\href
  {https://link.springer.com/article/10.1007/s11128-018-2002-y} {\bibfield
  {journal} {\bibinfo  {journal} {Quantum Information Processing}\ }\textbf
  {\bibinfo {volume} {17}},\ \bibinfo {pages} {1} (\bibinfo {year}
  {2018})}\BibitemShut {NoStop}%
\bibitem [{\citenamefont
  {Cervera-Lierta}(2018)}]{CerveraLierta2018exactisingmodel}%
  \BibitemOpen
  \bibfield  {author} {\bibinfo {author} {\bibfnamefont {A.}~\bibnamefont
  {Cervera-Lierta}},\ }\href {\doibase 10.22331/q-2018-12-21-114} {\bibfield
  {journal} {\bibinfo  {journal} {{Quantum}}\ }\textbf {\bibinfo {volume}
  {2}},\ \bibinfo {pages} {114} (\bibinfo {year} {2018})}\BibitemShut {NoStop}%
\bibitem [{\citenamefont {Francis}\ \emph {et~al.}(2020)\citenamefont
  {Francis}, \citenamefont {Freericks},\ and\ \citenamefont
  {Kemper}}]{francis2020quantum}%
  \BibitemOpen
  \bibfield  {author} {\bibinfo {author} {\bibfnamefont {A.}~\bibnamefont
  {Francis}}, \bibinfo {author} {\bibfnamefont {J.}~\bibnamefont {Freericks}},
  \ and\ \bibinfo {author} {\bibfnamefont {A.}~\bibnamefont {Kemper}},\ }\href
  {https://link.aps.org/doi/10.1103/PhysRevB.101.014411} {\bibfield  {journal}
  {\bibinfo  {journal} {Physical Review B}\ }\textbf {\bibinfo {volume}
  {101}},\ \bibinfo {pages} {014411} (\bibinfo {year} {2020})}\BibitemShut
  {NoStop}%
\bibitem [{\citenamefont {Smith}\ \emph {et~al.}(2019)\citenamefont {Smith},
  \citenamefont {Kim}, \citenamefont {Pollmann},\ and\ \citenamefont
  {Knolle}}]{smith2019simulating}%
  \BibitemOpen
  \bibfield  {author} {\bibinfo {author} {\bibfnamefont {A.}~\bibnamefont
  {Smith}}, \bibinfo {author} {\bibfnamefont {M.}~\bibnamefont {Kim}}, \bibinfo
  {author} {\bibfnamefont {F.}~\bibnamefont {Pollmann}}, \ and\ \bibinfo
  {author} {\bibfnamefont {J.}~\bibnamefont {Knolle}},\ }\href
  {https://www.nature.com/articles/s41534-019-0217-0} {\bibfield  {journal}
  {\bibinfo  {journal} {npj Quantum Information}\ }\textbf {\bibinfo {volume}
  {5}},\ \bibinfo {pages} {1} (\bibinfo {year} {2019})}\BibitemShut {NoStop}%
\bibitem [{\citenamefont {Vovrosh}\ and\ \citenamefont
  {Knolle}(2021)}]{vovrosh2021confinement}%
  \BibitemOpen
  \bibfield  {author} {\bibinfo {author} {\bibfnamefont {J.}~\bibnamefont
  {Vovrosh}}\ and\ \bibinfo {author} {\bibfnamefont {J.}~\bibnamefont
  {Knolle}},\ }\href {https://www.nature.com/articles/s41598-021-90849-5}
  {\bibfield  {journal} {\bibinfo  {journal} {Scientific reports}\ }\textbf
  {\bibinfo {volume} {11}},\ \bibinfo {pages} {1} (\bibinfo {year}
  {2021})}\BibitemShut {NoStop}%
\bibitem [{\citenamefont {Bulchandani}\ \emph {et~al.}(2021)\citenamefont
  {Bulchandani}, \citenamefont {Gopalakrishnan},\ and\ \citenamefont
  {Ilievski}}]{bulchandani2021superdiffusion}%
  \BibitemOpen
  \bibfield  {author} {\bibinfo {author} {\bibfnamefont {V.~B.}\ \bibnamefont
  {Bulchandani}}, \bibinfo {author} {\bibfnamefont {S.}~\bibnamefont
  {Gopalakrishnan}}, \ and\ \bibinfo {author} {\bibfnamefont {E.}~\bibnamefont
  {Ilievski}},\ }\href
  {https://iopscience.iop.org/article/10.1088/1742-5468/ac12c7/meta} {\bibfield
   {journal} {\bibinfo  {journal} {Journal of Statistical Mechanics: Theory and
  Experiment}\ }\textbf {\bibinfo {volume} {2021}},\ \bibinfo {pages} {084001}
  (\bibinfo {year} {2021})}\BibitemShut {NoStop}%
\bibitem [{\citenamefont {Fourier}(1822)}]{fourier1822theorie}%
  \BibitemOpen
  \bibfield  {author} {\bibinfo {author} {\bibfnamefont {J.}~\bibnamefont
  {Fourier}},\ }\href
  {https://www.cambridge.org/core/books/theorie-analytique-de-la-chaleur/6A2F7B17FC2BCBDA255300D171C95B34}
  {\enquote {\bibinfo {title} {Th{\'e}orie analytique de la chaleur},}\ }
  (\bibinfo {year} {1822})\BibitemShut {NoStop}%
\bibitem [{\citenamefont {Buchanan}(2005)}]{buchanan2005heated}%
  \BibitemOpen
  \bibfield  {author} {\bibinfo {author} {\bibfnamefont {M.}~\bibnamefont
  {Buchanan}},\ }\href {https://www.nature.com/articles/nphys157} {\bibfield
  {journal} {\bibinfo  {journal} {Nature Physics}\ }\textbf {\bibinfo {volume}
  {1}},\ \bibinfo {pages} {71} (\bibinfo {year} {2005})}\BibitemShut {NoStop}%
\bibitem [{\citenamefont {Bertini}\ \emph {et~al.}(2021)\citenamefont
  {Bertini}, \citenamefont {Heidrich-Meisner}, \citenamefont {Karrasch},
  \citenamefont {Prosen}, \citenamefont {Steinigeweg},\ and\ \citenamefont
  {\ifmmode \check{Z}\else \v{Z}\fi{}nidari\ifmmode~\check{c}\else
  \v{c}\fi{}}}]{bertini_review}%
  \BibitemOpen
  \bibfield  {author} {\bibinfo {author} {\bibfnamefont {B.}~\bibnamefont
  {Bertini}}, \bibinfo {author} {\bibfnamefont {F.}~\bibnamefont
  {Heidrich-Meisner}}, \bibinfo {author} {\bibfnamefont {C.}~\bibnamefont
  {Karrasch}}, \bibinfo {author} {\bibfnamefont {T.}~\bibnamefont {Prosen}},
  \bibinfo {author} {\bibfnamefont {R.}~\bibnamefont {Steinigeweg}}, \ and\
  \bibinfo {author} {\bibfnamefont {M.}~\bibnamefont {\ifmmode \check{Z}\else
  \v{Z}\fi{}nidari\ifmmode~\check{c}\else \v{c}\fi{}}},\ }\href {\doibase
  10.1103/RevModPhys.93.025003} {\bibfield  {journal} {\bibinfo  {journal}
  {Rev. Mod. Phys.}\ }\textbf {\bibinfo {volume} {93}},\ \bibinfo {pages}
  {025003} (\bibinfo {year} {2021})}\BibitemShut {NoStop}%
\bibitem [{\citenamefont {\ifmmode \check{Z}\else
  \v{Z}\fi{}nidari\ifmmode~\check{c}\else
  \v{c}\fi{}}(2011)}]{Marko_boundary_driving_XXZ}%
  \BibitemOpen
  \bibfield  {author} {\bibinfo {author} {\bibfnamefont {M.}~\bibnamefont
  {\ifmmode \check{Z}\else \v{Z}\fi{}nidari\ifmmode~\check{c}\else
  \v{c}\fi{}}},\ }\href {\doibase 10.1103/PhysRevLett.106.220601} {\bibfield
  {journal} {\bibinfo  {journal} {Phys. Rev. Lett.}\ }\textbf {\bibinfo
  {volume} {106}},\ \bibinfo {pages} {220601} (\bibinfo {year}
  {2011})}\BibitemShut {NoStop}%
\bibitem [{\citenamefont {Ljubotina}\ \emph
  {et~al.}(2019{\natexlab{a}})\citenamefont {Ljubotina}, \citenamefont
  {\ifmmode \check{Z}\else \v{Z}\fi{}nidari\ifmmode~\check{c}\else
  \v{c}\fi{}},\ and\ \citenamefont {Prosen}}]{PhysRevLett.122.210602}%
  \BibitemOpen
  \bibfield  {author} {\bibinfo {author} {\bibfnamefont {M.}~\bibnamefont
  {Ljubotina}}, \bibinfo {author} {\bibfnamefont {M.}~\bibnamefont {\ifmmode
  \check{Z}\else \v{Z}\fi{}nidari\ifmmode~\check{c}\else \v{c}\fi{}}}, \ and\
  \bibinfo {author} {\bibfnamefont {T.}~\bibnamefont {Prosen}},\ }\href
  {\doibase 10.1103/PhysRevLett.122.210602} {\bibfield  {journal} {\bibinfo
  {journal} {Phys. Rev. Lett.}\ }\textbf {\bibinfo {volume} {122}},\ \bibinfo
  {pages} {210602} (\bibinfo {year} {2019}{\natexlab{a}})}\BibitemShut
  {NoStop}%
\bibitem [{\citenamefont {Kardar}\ \emph {et~al.}(1986)\citenamefont {Kardar},
  \citenamefont {Parisi},\ and\ \citenamefont {Zhang}}]{kpz_original}%
  \BibitemOpen
  \bibfield  {author} {\bibinfo {author} {\bibfnamefont {M.}~\bibnamefont
  {Kardar}}, \bibinfo {author} {\bibfnamefont {G.}~\bibnamefont {Parisi}}, \
  and\ \bibinfo {author} {\bibfnamefont {Y.-C.}\ \bibnamefont {Zhang}},\ }\href
  {\doibase 10.1103/PhysRevLett.56.889} {\bibfield  {journal} {\bibinfo
  {journal} {Phys. Rev. Lett.}\ }\textbf {\bibinfo {volume} {56}},\ \bibinfo
  {pages} {889} (\bibinfo {year} {1986})}\BibitemShut {NoStop}%
\bibitem [{\citenamefont {Dupont}\ \emph {et~al.}(2021)\citenamefont {Dupont},
  \citenamefont {Sherman},\ and\ \citenamefont
  {Moore}}]{PhysRevLett.127.107201}%
  \BibitemOpen
  \bibfield  {author} {\bibinfo {author} {\bibfnamefont {M.}~\bibnamefont
  {Dupont}}, \bibinfo {author} {\bibfnamefont {N.~E.}\ \bibnamefont {Sherman}},
  \ and\ \bibinfo {author} {\bibfnamefont {J.~E.}\ \bibnamefont {Moore}},\
  }\href {\doibase 10.1103/PhysRevLett.127.107201} {\bibfield  {journal}
  {\bibinfo  {journal} {Phys. Rev. Lett.}\ }\textbf {\bibinfo {volume} {127}},\
  \bibinfo {pages} {107201} (\bibinfo {year} {2021})}\BibitemShut {NoStop}%
\bibitem [{\citenamefont {Ilievski}\ \emph {et~al.}(2021)\citenamefont
  {Ilievski}, \citenamefont {De~Nardis}, \citenamefont {Gopalakrishnan},
  \citenamefont {Vasseur},\ and\ \citenamefont {Ware}}]{PhysRevX.11.031023}%
  \BibitemOpen
  \bibfield  {author} {\bibinfo {author} {\bibfnamefont {E.}~\bibnamefont
  {Ilievski}}, \bibinfo {author} {\bibfnamefont {J.}~\bibnamefont {De~Nardis}},
  \bibinfo {author} {\bibfnamefont {S.}~\bibnamefont {Gopalakrishnan}},
  \bibinfo {author} {\bibfnamefont {R.}~\bibnamefont {Vasseur}}, \ and\
  \bibinfo {author} {\bibfnamefont {B.}~\bibnamefont {Ware}},\ }\href {\doibase
  10.1103/PhysRevX.11.031023} {\bibfield  {journal} {\bibinfo  {journal} {Phys.
  Rev. X}\ }\textbf {\bibinfo {volume} {11}},\ \bibinfo {pages} {031023}
  (\bibinfo {year} {2021})}\BibitemShut {NoStop}%
\bibitem [{\citenamefont {Gopalakrishnan}\ and\ \citenamefont
  {Vasseur}(2022)}]{sarang}%
  \BibitemOpen
  \bibfield  {author} {\bibinfo {author} {\bibfnamefont {S.}~\bibnamefont
  {Gopalakrishnan}}\ and\ \bibinfo {author} {\bibfnamefont {R.}~\bibnamefont
  {Vasseur}},\ }\href {https://arxiv.org/abs/2208.11133} {\bibfield  {journal}
  {\bibinfo  {journal} {arXiv preprint arXiv:2208.11133}\ } (\bibinfo {year}
  {2022})}\BibitemShut {NoStop}%
\bibitem [{\citenamefont {Scheie}\ \emph {et~al.}(2021)\citenamefont {Scheie},
  \citenamefont {Sherman}, \citenamefont {Dupont}, \citenamefont {Nagler},
  \citenamefont {Stone}, \citenamefont {Granroth}, \citenamefont {Moore},\ and\
  \citenamefont {Tennant}}]{scheie2021detection}%
  \BibitemOpen
  \bibfield  {author} {\bibinfo {author} {\bibfnamefont {A.}~\bibnamefont
  {Scheie}}, \bibinfo {author} {\bibfnamefont {N.}~\bibnamefont {Sherman}},
  \bibinfo {author} {\bibfnamefont {M.}~\bibnamefont {Dupont}}, \bibinfo
  {author} {\bibfnamefont {S.}~\bibnamefont {Nagler}}, \bibinfo {author}
  {\bibfnamefont {M.}~\bibnamefont {Stone}}, \bibinfo {author} {\bibfnamefont
  {G.}~\bibnamefont {Granroth}}, \bibinfo {author} {\bibfnamefont
  {J.}~\bibnamefont {Moore}}, \ and\ \bibinfo {author} {\bibfnamefont
  {D.}~\bibnamefont {Tennant}},\ }\href
  {https://www.nature.com/articles/s41567-021-01191-6} {\bibfield  {journal}
  {\bibinfo  {journal} {Nature Physics}\ }\textbf {\bibinfo {volume} {17}},\
  \bibinfo {pages} {726} (\bibinfo {year} {2021})}\BibitemShut {NoStop}%
\bibitem [{\citenamefont {Wei}\ \emph {et~al.}(2022)\citenamefont {Wei},
  \citenamefont {Rubio-Abadal}, \citenamefont {Ye}, \citenamefont {Machado},
  \citenamefont {Kemp}, \citenamefont {Srakaew}, \citenamefont {Hollerith},
  \citenamefont {Rui}, \citenamefont {Gopalakrishnan}, \citenamefont {Yao}
  \emph {et~al.}}]{wei2022quantum}%
  \BibitemOpen
  \bibfield  {author} {\bibinfo {author} {\bibfnamefont {D.}~\bibnamefont
  {Wei}}, \bibinfo {author} {\bibfnamefont {A.}~\bibnamefont {Rubio-Abadal}},
  \bibinfo {author} {\bibfnamefont {B.}~\bibnamefont {Ye}}, \bibinfo {author}
  {\bibfnamefont {F.}~\bibnamefont {Machado}}, \bibinfo {author} {\bibfnamefont
  {J.}~\bibnamefont {Kemp}}, \bibinfo {author} {\bibfnamefont {K.}~\bibnamefont
  {Srakaew}}, \bibinfo {author} {\bibfnamefont {S.}~\bibnamefont {Hollerith}},
  \bibinfo {author} {\bibfnamefont {J.}~\bibnamefont {Rui}}, \bibinfo {author}
  {\bibfnamefont {S.}~\bibnamefont {Gopalakrishnan}}, \bibinfo {author}
  {\bibfnamefont {N.~Y.}\ \bibnamefont {Yao}},  \emph {et~al.},\ }\href
  {https://www.science.org/doi/abs/10.1126/science.abk2397?casa_token=-z7ymyTkxZQAAAAA:yf_J3zwpcQplB6yiyqn34ZoC3_z4EdZYhikjfogQ6Wr9HEw7IYYtNX_64MYxjaoBeSh09YbU75rALrc}
  {\bibfield  {journal} {\bibinfo  {journal} {Science}\ }\textbf {\bibinfo
  {volume} {376}},\ \bibinfo {pages} {716} (\bibinfo {year}
  {2022})}\BibitemShut {NoStop}%
\bibitem [{\citenamefont {Joshi}\ \emph {et~al.}(2022)\citenamefont {Joshi},
  \citenamefont {Kranzl}, \citenamefont {Schuckert}, \citenamefont {Lovas},
  \citenamefont {Maier}, \citenamefont {Blatt}, \citenamefont {Knap},\ and\
  \citenamefont {Roos}}]{joshi2022observing}%
  \BibitemOpen
  \bibfield  {author} {\bibinfo {author} {\bibfnamefont {M.~K.}\ \bibnamefont
  {Joshi}}, \bibinfo {author} {\bibfnamefont {F.}~\bibnamefont {Kranzl}},
  \bibinfo {author} {\bibfnamefont {A.}~\bibnamefont {Schuckert}}, \bibinfo
  {author} {\bibfnamefont {I.}~\bibnamefont {Lovas}}, \bibinfo {author}
  {\bibfnamefont {C.}~\bibnamefont {Maier}}, \bibinfo {author} {\bibfnamefont
  {R.}~\bibnamefont {Blatt}}, \bibinfo {author} {\bibfnamefont
  {M.}~\bibnamefont {Knap}}, \ and\ \bibinfo {author} {\bibfnamefont {C.~F.}\
  \bibnamefont {Roos}},\ }\href
  {https://www.science.org/doi/10.1126/science.abk2400} {\bibfield  {journal}
  {\bibinfo  {journal} {Science}\ }\textbf {\bibinfo {volume} {376}},\ \bibinfo
  {pages} {720} (\bibinfo {year} {2022})}\BibitemShut {NoStop}%
\bibitem [{\citenamefont {Fontaine}\ \emph {et~al.}(2022)\citenamefont
  {Fontaine} \emph {et~al.}}]{polariton}%
  \BibitemOpen
  \bibfield  {author} {\bibinfo {author} {\bibfnamefont {Q.}~\bibnamefont
  {Fontaine}} \emph {et~al.},\ }\href
  {https://www.nature.com/articles/s41586-022-05001-8} {\bibfield  {journal}
  {\bibinfo  {journal} {Nature}\ }\textbf {\bibinfo {volume} {608}},\ \bibinfo
  {pages} {687} (\bibinfo {year} {2022})}\BibitemShut {NoStop}%
\bibitem [{\citenamefont {Vanicat}\ \emph {et~al.}(2018)\citenamefont
  {Vanicat}, \citenamefont {Zadnik},\ and\ \citenamefont
  {Prosen}}]{PhysRevLett.121.030606}%
  \BibitemOpen
  \bibfield  {author} {\bibinfo {author} {\bibfnamefont {M.}~\bibnamefont
  {Vanicat}}, \bibinfo {author} {\bibfnamefont {L.}~\bibnamefont {Zadnik}}, \
  and\ \bibinfo {author} {\bibfnamefont {T.}~\bibnamefont {Prosen}},\ }\href
  {\doibase 10.1103/PhysRevLett.121.030606} {\bibfield  {journal} {\bibinfo
  {journal} {Phys. Rev. Lett.}\ }\textbf {\bibinfo {volume} {121}},\ \bibinfo
  {pages} {030606} (\bibinfo {year} {2018})}\BibitemShut {NoStop}%
\bibitem [{\citenamefont {Krajnik}\ and\ \citenamefont
  {Prosen}(2020)}]{krajnik2020kardar}%
  \BibitemOpen
  \bibfield  {author} {\bibinfo {author} {\bibfnamefont {{\v{Z}}.}~\bibnamefont
  {Krajnik}}\ and\ \bibinfo {author} {\bibfnamefont {T.}~\bibnamefont
  {Prosen}},\ }\href {https://doi.org/10.1007/s10955-020-02523-1} {\bibfield
  {journal} {\bibinfo  {journal} {Journal of Statistical Physics}\ }\textbf
  {\bibinfo {volume} {179}},\ \bibinfo {pages} {110} (\bibinfo {year}
  {2020})}\BibitemShut {NoStop}%
\bibitem [{\citenamefont {Richter}\ and\ \citenamefont
  {Pal}(2021)}]{PhysRevLett.126.230501}%
  \BibitemOpen
  \bibfield  {author} {\bibinfo {author} {\bibfnamefont {J.}~\bibnamefont
  {Richter}}\ and\ \bibinfo {author} {\bibfnamefont {A.}~\bibnamefont {Pal}},\
  }\href {\doibase 10.1103/PhysRevLett.126.230501} {\bibfield  {journal}
  {\bibinfo  {journal} {Phys. Rev. Lett.}\ }\textbf {\bibinfo {volume} {126}},\
  \bibinfo {pages} {230501} (\bibinfo {year} {2021})}\BibitemShut {NoStop}%
\bibitem [{\citenamefont {Lunt}\ \emph {et~al.}(2021)\citenamefont {Lunt},
  \citenamefont {Richter},\ and\ \citenamefont {Pal}}]{lunt2021quantum}%
  \BibitemOpen
  \bibfield  {author} {\bibinfo {author} {\bibfnamefont {O.}~\bibnamefont
  {Lunt}}, \bibinfo {author} {\bibfnamefont {J.}~\bibnamefont {Richter}}, \
  and\ \bibinfo {author} {\bibfnamefont {A.}~\bibnamefont {Pal}},\ }\href
  {https://arxiv.org/abs/2112.06682} {\bibfield  {journal} {\bibinfo  {journal}
  {arXiv preprint arXiv:2112.06682}\ } (\bibinfo {year} {2021})}\BibitemShut
  {NoStop}%
\bibitem [{\citenamefont {Jin}\ \emph {et~al.}(2021)\citenamefont {Jin},
  \citenamefont {Willsch}, \citenamefont {Willsch}, \citenamefont {Lagemann},
  \citenamefont {Michielsen},\ and\ \citenamefont {De~Raedt}}]{jin2021random}%
  \BibitemOpen
  \bibfield  {author} {\bibinfo {author} {\bibfnamefont {F.}~\bibnamefont
  {Jin}}, \bibinfo {author} {\bibfnamefont {D.}~\bibnamefont {Willsch}},
  \bibinfo {author} {\bibfnamefont {M.}~\bibnamefont {Willsch}}, \bibinfo
  {author} {\bibfnamefont {H.}~\bibnamefont {Lagemann}}, \bibinfo {author}
  {\bibfnamefont {K.}~\bibnamefont {Michielsen}}, \ and\ \bibinfo {author}
  {\bibfnamefont {H.}~\bibnamefont {De~Raedt}},\ }\href
  {https://journals.jps.jp/doi/full/10.7566/JPSJ.90.012001} {\bibfield
  {journal} {\bibinfo  {journal} {Journal of the Physical Society of Japan}\
  }\textbf {\bibinfo {volume} {90}},\ \bibinfo {pages} {012001} (\bibinfo
  {year} {2021})}\BibitemShut {NoStop}%
\bibitem [{\citenamefont {Arute}\ \emph {et~al.}(2019)\citenamefont {Arute},
  \citenamefont {Arya}, \citenamefont {Babbush}, \citenamefont {Bacon},
  \citenamefont {Bardin}, \citenamefont {Barends}, \citenamefont {Biswas},
  \citenamefont {Boixo}, \citenamefont {Brandao}, \citenamefont {Buell} \emph
  {et~al.}}]{arute2019quantum}%
  \BibitemOpen
  \bibfield  {author} {\bibinfo {author} {\bibfnamefont {F.}~\bibnamefont
  {Arute}}, \bibinfo {author} {\bibfnamefont {K.}~\bibnamefont {Arya}},
  \bibinfo {author} {\bibfnamefont {R.}~\bibnamefont {Babbush}}, \bibinfo
  {author} {\bibfnamefont {D.}~\bibnamefont {Bacon}}, \bibinfo {author}
  {\bibfnamefont {J.~C.}\ \bibnamefont {Bardin}}, \bibinfo {author}
  {\bibfnamefont {R.}~\bibnamefont {Barends}}, \bibinfo {author} {\bibfnamefont
  {R.}~\bibnamefont {Biswas}}, \bibinfo {author} {\bibfnamefont
  {S.}~\bibnamefont {Boixo}}, \bibinfo {author} {\bibfnamefont {F.~G.}\
  \bibnamefont {Brandao}}, \bibinfo {author} {\bibfnamefont {D.~A.}\
  \bibnamefont {Buell}},  \emph {et~al.},\ }\href
  {https://www.nature.com/articles/s41586-019-1666-5} {\bibfield  {journal}
  {\bibinfo  {journal} {Nature}\ }\textbf {\bibinfo {volume} {574}},\ \bibinfo
  {pages} {505} (\bibinfo {year} {2019})}\BibitemShut {NoStop}%
\bibitem [{\citenamefont {Page}(1993)}]{page_93}%
  \BibitemOpen
  \bibfield  {author} {\bibinfo {author} {\bibfnamefont {D.~N.}\ \bibnamefont
  {Page}},\ }\href {\doibase 10.1103/PhysRevLett.71.3743} {\bibfield  {journal}
  {\bibinfo  {journal} {Phys. Rev. Lett.}\ }\textbf {\bibinfo {volume} {71}},\
  \bibinfo {pages} {3743} (\bibinfo {year} {1993})}\BibitemShut {NoStop}%
\bibitem [{ibm()}]{ibm}%
  \BibitemOpen
  \href
  {https://quantum-computing.ibm.com/services/resources?tab=systems&system=ibmq_montreal}
  {\enquote {\bibinfo {title} {Ibm quantum compute resources},}\ }\BibitemShut
  {NoStop}%
\bibitem [{\citenamefont {Steinigeweg}\ \emph {et~al.}(2014)\citenamefont
  {Steinigeweg}, \citenamefont {Gemmer},\ and\ \citenamefont
  {Brenig}}]{PhysRevLett.112.120601}%
  \BibitemOpen
  \bibfield  {author} {\bibinfo {author} {\bibfnamefont {R.}~\bibnamefont
  {Steinigeweg}}, \bibinfo {author} {\bibfnamefont {J.}~\bibnamefont {Gemmer}},
  \ and\ \bibinfo {author} {\bibfnamefont {W.}~\bibnamefont {Brenig}},\ }\href
  {\doibase 10.1103/PhysRevLett.112.120601} {\bibfield  {journal} {\bibinfo
  {journal} {Phys. Rev. Lett.}\ }\textbf {\bibinfo {volume} {112}},\ \bibinfo
  {pages} {120601} (\bibinfo {year} {2014})}\BibitemShut {NoStop}%
\bibitem [{\citenamefont {Richter}\ \emph {et~al.}(2019)\citenamefont
  {Richter}, \citenamefont {Jin}, \citenamefont {Knipschild}, \citenamefont
  {Herbrych}, \citenamefont {De~Raedt}, \citenamefont {Michielsen},
  \citenamefont {Gemmer},\ and\ \citenamefont
  {Steinigeweg}}]{PhysRevB.99.144422}%
  \BibitemOpen
  \bibfield  {author} {\bibinfo {author} {\bibfnamefont {J.}~\bibnamefont
  {Richter}}, \bibinfo {author} {\bibfnamefont {F.}~\bibnamefont {Jin}},
  \bibinfo {author} {\bibfnamefont {L.}~\bibnamefont {Knipschild}}, \bibinfo
  {author} {\bibfnamefont {J.}~\bibnamefont {Herbrych}}, \bibinfo {author}
  {\bibfnamefont {H.}~\bibnamefont {De~Raedt}}, \bibinfo {author}
  {\bibfnamefont {K.}~\bibnamefont {Michielsen}}, \bibinfo {author}
  {\bibfnamefont {J.}~\bibnamefont {Gemmer}}, \ and\ \bibinfo {author}
  {\bibfnamefont {R.}~\bibnamefont {Steinigeweg}},\ }\href {\doibase
  10.1103/PhysRevB.99.144422} {\bibfield  {journal} {\bibinfo  {journal} {Phys.
  Rev. B}\ }\textbf {\bibinfo {volume} {99}},\ \bibinfo {pages} {144422}
  (\bibinfo {year} {2019})}\BibitemShut {NoStop}%
\bibitem [{\citenamefont {Richter}(2020)}]{richter2020quantum}%
  \BibitemOpen
  \bibfield  {author} {\bibinfo {author} {\bibfnamefont {J.}~\bibnamefont
  {Richter}},\ }\href
  {https://inis.iaea.org/search/search.aspx?orig_q=RN:52016158} {\bibfield
  {journal} {\bibinfo  {journal} {Universit{\"a}t Osnabr{\"u}ck}\ } (\bibinfo
  {year} {2020})}\BibitemShut {NoStop}%
\bibitem [{\citenamefont {Chiaracane}\ \emph {et~al.}(2021)\citenamefont
  {Chiaracane}, \citenamefont {Pietracaprina}, \citenamefont {Purkayastha},\
  and\ \citenamefont {Goold}}]{PhysRevB.103.184205}%
  \BibitemOpen
  \bibfield  {author} {\bibinfo {author} {\bibfnamefont {C.}~\bibnamefont
  {Chiaracane}}, \bibinfo {author} {\bibfnamefont {F.}~\bibnamefont
  {Pietracaprina}}, \bibinfo {author} {\bibfnamefont {A.}~\bibnamefont
  {Purkayastha}}, \ and\ \bibinfo {author} {\bibfnamefont {J.}~\bibnamefont
  {Goold}},\ }\href {\doibase 10.1103/PhysRevB.103.184205} {\bibfield
  {journal} {\bibinfo  {journal} {Phys. Rev. B}\ }\textbf {\bibinfo {volume}
  {103}},\ \bibinfo {pages} {184205} (\bibinfo {year} {2021})}\BibitemShut
  {NoStop}%
\bibitem [{\citenamefont {Vatan}\ and\ \citenamefont
  {Williams}(2004)}]{williams}%
  \BibitemOpen
  \bibfield  {author} {\bibinfo {author} {\bibfnamefont {F.}~\bibnamefont
  {Vatan}}\ and\ \bibinfo {author} {\bibfnamefont {C.}~\bibnamefont
  {Williams}},\ }\href {\doibase 10.1103/PhysRevA.69.032315} {\bibfield
  {journal} {\bibinfo  {journal} {Phys. Rev. A}\ }\textbf {\bibinfo {volume}
  {69}},\ \bibinfo {pages} {032315} (\bibinfo {year} {2004})}\BibitemShut
  {NoStop}%
\bibitem [{\citenamefont {Maruyoshi}\ \emph {et~al.}(2022)\citenamefont
  {Maruyoshi} \emph {et~al.}}]{trotter_quantum}%
  \BibitemOpen
  \bibfield  {author} {\bibinfo {author} {\bibfnamefont {K.}~\bibnamefont
  {Maruyoshi}} \emph {et~al.},\ }\href {https://arxiv.org/abs/2208.00576}
  {\bibfield  {journal} {\bibinfo  {journal} {arXiv preprint arXiv:2208.00576}\
  } (\bibinfo {year} {2022})}\BibitemShut {NoStop}%
\bibitem [{sm()}]{sm}%
  \BibitemOpen
  \href@noop {} {}\bibinfo {note} {See Supplemental Material.}\BibitemShut
  {Stop}%
\bibitem [{\citenamefont {Geller}\ \emph {et~al.}(2022)\citenamefont {Geller},
  \citenamefont {Arrasmith}, \citenamefont {Holmes}, \citenamefont {Yan},
  \citenamefont {Coles},\ and\ \citenamefont {Sornborger}}]{weave}%
  \BibitemOpen
  \bibfield  {author} {\bibinfo {author} {\bibfnamefont {M.~R.}\ \bibnamefont
  {Geller}}, \bibinfo {author} {\bibfnamefont {A.}~\bibnamefont {Arrasmith}},
  \bibinfo {author} {\bibfnamefont {Z.}~\bibnamefont {Holmes}}, \bibinfo
  {author} {\bibfnamefont {B.}~\bibnamefont {Yan}}, \bibinfo {author}
  {\bibfnamefont {P.~J.}\ \bibnamefont {Coles}}, \ and\ \bibinfo {author}
  {\bibfnamefont {A.}~\bibnamefont {Sornborger}},\ }\href {\doibase
  10.1103/PhysRevE.105.035302} {\bibfield  {journal} {\bibinfo  {journal}
  {Phys. Rev. E}\ }\textbf {\bibinfo {volume} {105}},\ \bibinfo {pages}
  {035302} (\bibinfo {year} {2022})}\BibitemShut {NoStop}%
\bibitem [{\citenamefont {Li}\ and\ \citenamefont
  {Benjamin}(2017)}]{li2017efficient}%
  \BibitemOpen
  \bibfield  {author} {\bibinfo {author} {\bibfnamefont {Y.}~\bibnamefont
  {Li}}\ and\ \bibinfo {author} {\bibfnamefont {S.~C.}\ \bibnamefont
  {Benjamin}},\ }\href@noop {} {\bibfield  {journal} {\bibinfo  {journal}
  {Physical Review X}\ }\textbf {\bibinfo {volume} {7}},\ \bibinfo {pages}
  {021050} (\bibinfo {year} {2017})}\BibitemShut {NoStop}%
\bibitem [{\citenamefont {Giurgica-Tiron}\ \emph {et~al.}(2020)\citenamefont
  {Giurgica-Tiron}, \citenamefont {Hindy}, \citenamefont {LaRose},
  \citenamefont {Mari},\ and\ \citenamefont {Zeng}}]{giurgica2020digital}%
  \BibitemOpen
  \bibfield  {author} {\bibinfo {author} {\bibfnamefont {T.}~\bibnamefont
  {Giurgica-Tiron}}, \bibinfo {author} {\bibfnamefont {Y.}~\bibnamefont
  {Hindy}}, \bibinfo {author} {\bibfnamefont {R.}~\bibnamefont {LaRose}},
  \bibinfo {author} {\bibfnamefont {A.}~\bibnamefont {Mari}}, \ and\ \bibinfo
  {author} {\bibfnamefont {W.~J.}\ \bibnamefont {Zeng}},\ }in\ \href@noop {}
  {\emph {\bibinfo {booktitle} {2020 IEEE International Conference on Quantum
  Computing and Engineering (QCE)}}}\ (\bibinfo {organization} {IEEE},\
  \bibinfo {year} {2020})\ pp.\ \bibinfo {pages} {306--316}\BibitemShut
  {NoStop}%
\bibitem [{\citenamefont {Ljubotina}\ \emph
  {et~al.}(2019{\natexlab{b}})\citenamefont {Ljubotina}, \citenamefont
  {Zadnik},\ and\ \citenamefont {Prosen}}]{ljubotina2019ballistic}%
  \BibitemOpen
  \bibfield  {author} {\bibinfo {author} {\bibfnamefont {M.}~\bibnamefont
  {Ljubotina}}, \bibinfo {author} {\bibfnamefont {L.}~\bibnamefont {Zadnik}}, \
  and\ \bibinfo {author} {\bibfnamefont {T.}~\bibnamefont {Prosen}},\
  }\href@noop {} {\bibfield  {journal} {\bibinfo  {journal} {Physical review
  letters}\ }\textbf {\bibinfo {volume} {122}},\ \bibinfo {pages} {150605}
  (\bibinfo {year} {2019}{\natexlab{b}})}\BibitemShut {NoStop}%
\end{thebibliography}%

\end{document}